\documentclass[preprint,12pt,3p]{elsarticle}
\usepackage[bookmarks=false]{hyperref}
\usepackage{float}
\usepackage{setspace}
\usepackage{lineno}
\usepackage{amsmath}
\usepackage{makecell}
\usepackage{multirow}

\begin{document}
\title{Ground calibration of Gamma-Ray Detectors of GECAM-C}

\cortext[cor1]{Corresponding author.}
\author[1,2]{Chao Zheng}
\author[1]{Zheng-Hua An\corref{cor1}}
\ead{anzh@ihep.ac.cn}
\author[1]{Wen-Xi Peng\corref{cor1}}
\ead{pengwx@ihep.ac.cn}
\author[1]{Da-Li Zhang\corref{cor1}}
\ead{zhangdl@ihep.ac.cn}
\author[1]{Shao-Lin Xiong}
\author[1]{Rui. Qiao}
\author[1,2]{Yan-Qiu Zhang}
\author[1,2]{Wang-Chen Xue}
\author[1,2]{Jia-Cong Liu}
\author[1,2]{Pei-Yi Feng}
\author[1,2,6]{Ce. Cai}
\author[1]{Min Gao}
\author[1]{Ke Gong}
\author[1]{Dong-Ya Guo}
\author[1]{Dong-Jie Hou}
\author[1]{Gang Li}
\author[1]{Xin-Qiao Li}
\author[1]{Yan-Guo Li}
\author[1]{Mao-Shun Li}
\author[1]{Xiao-Hua Liang}
\author[1]{Ya-Qing Liu}
\author[1]{Xiao-Jing Liu}
\author[1]{Li-Ming Song}
\author[1]{Xi-Lei Sun}
\author[1,2]{Wen-Jun Tan}
\author[1,2]{Chen-Wei Wang}
\author[1]{Hui Wang}
\author[1]{Jin-Zhou Wang}
\author[1]{Xiang-Yang Wen}
\author[1,2,5]{Shuo Xiao}
\author[1]{Yan-Bing Xu}
\author[1]{Sheng Yang}
\author[1,3]{Qi-Bing Yi}
\author[1]{Fan Zhang}
\author[1]{Peng Zhang}
\author[1]{Zhen Zhang}
\author[1,3]{Yi Zhao}
\author[1,2]{Xing Zhou}

\address[1]{Key Laboratory of Particle Astrophysics, Institute of High Energy Physics,
Chinese Academy of Sciences, Beijing 100049, China}
\address[2]{University of Chinese Academy of Sciences, Chinese Academy of Sciences, Beijing 100049, China}
\address[3]{Key Laboratory of Stellar and Interstellar Physics and Department of Physics, Xiangtan University, Xiangtan 411105, China}
\address[4]{Department of Astronomy, Beijing Normal University, Beijing 100875, China}
\address[5]{School of Physics and Electronic Science, Guizhou Normal University, Guiyang 550001, People’s Republic of China}
\address[6]{College of Physics and Hebei Key Laboratory of Photophysics Research and Application, Hebei Normal University, Shijiazhuang, Hebei 050024, China}

\begin{abstract}
As a new member of GECAM mission, GECAM-C (also named High Energy Burst Searcher, HEBS) was launched onboard the SATech-01 satellite on July 27th, 2022, which is capable to monitor gamma-ray transients from $\sim$ 6 keV to 6 MeV. As the main detector, there are 12 gamma-ray detectors (GRDs) equipped for GECAM-C.
In order to verify the GECAM-C GRD detector performance and to validate the Monte Carlo simulations of detector response, comprehensive on-ground calibration experiments have been performed using X-ray beam and radioactive sources, including Energy–Channel relation, energy resolution, detection efficiency, SiPM voltage-gain relation and the non-uniformity of positional response. In this paper, the detailed calibration campaigns and data analysis results for GECAM-C GRDs are presented, demonstrating the excellent performance of GECAM-C GRD detectors.

\end{abstract}

\begin{keyword}
Calibration $\cdot$ LaBr$_3$(Ce/Ce+Sr) detector $\cdot$ NaI (Tl)detector $\cdot$ Gamma-Ray Burst $\cdot$ GECAM-C
\end{keyword}

\maketitle

\section{Introduction}
On August 17, 2017, the LIGO and Virgo discovered a gravitational wave (GW170817) produced by binary neutron star merger whose electromagnetic counterpart were observed by multi-wavelength telescopes around the world, heralding the era of multi-messenger gravitational wave astronomy\cite{GW170817,GRB170817,Multi_messenger}. In the observation campaign of GW170817, the high-energy electromagnetic counterpart, GRB 170817A, played an important role. 

\begin{figure}[H]
\centering 
\includegraphics[width=16cm,height=5cm]{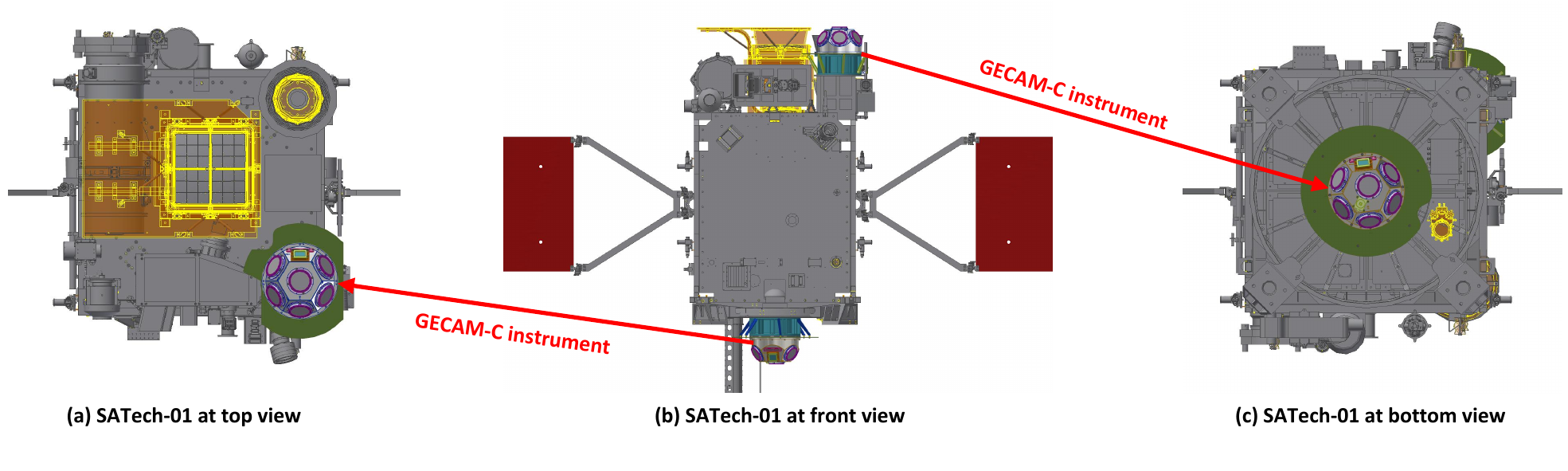}
\caption{GECAM-C (i.e. HEBS) onboard the SATech-01 satellite. GECAM-C is composed of two detector domes: top dome and bottom dome.}
\label{SY-01_sat}
\end{figure}

GECAM\cite{GECAM,xiao2022ground,xiao2021} is a dedicated mission to observe gamma-ray transients, especially those GRBs associated with GW events. Originally, GECAM mission is composed of two micro-satellites, i.e. GECAM-A and GECAM-B, which have been launched in December 2020. 
As the third member of GECAM mission, the GECAM-C (also named High Energy Burst Searcher, HEBS) is designed to improve the detection capability of GECAM mission. GECAM-C was launched onboard SATech-01 satellite\footnote{ \href{https://www.globaltimes.cn/page/202301/1283692.shtml}{https://www.globaltimes.cn/page/202301/1283692.shtml}} on July 27th, 2022. GECAM-C operates in the sun-synchronous orbit with an altitude about 500 km and an inclination of 97.4° (see Fig.\ref{SY-01_sat}).  

As the main detector of GECAM-C, Gamma-Ray Detector (GRD) is designed to measure positional, temporal and spectral properties of gamma-ray transients, including gravitational wave gamma-ray bursts, fast radio bursts and other high energy transient sources \cite{objective0,objective2,objective3}. GRDs are made of LaBr$_3$ (Ce/Ce+Sr) or NaI(Tl) scintillators readout by SiPM array.
The main characteristics of GRD are listed in Table.\ref{GRD designed}.

\begin{figure}[H]
\centering
\includegraphics[width=12cm,height=6cm]{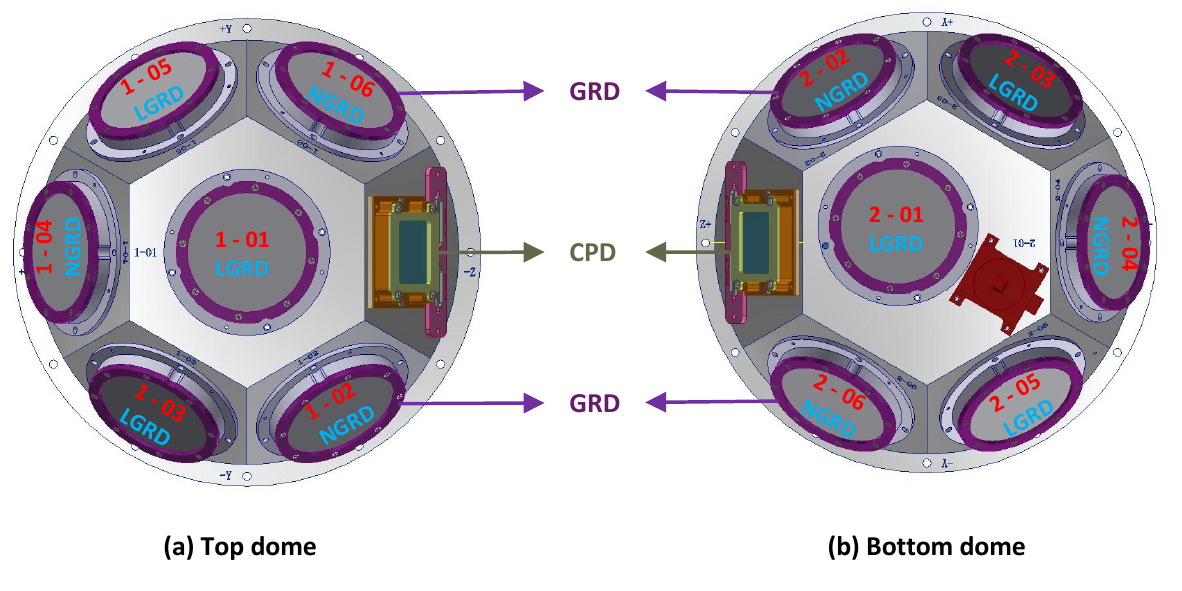}
\caption{Detector domes of GECAM-C. Each dome is equipped with different types of GRDs, including NaI-based NGRD and LaBr$_3$-based LGRD.}
\label{dome}
\end{figure}

In order to achieve large field of view and to do localization of the gamma-ray transients \cite{Liao2020, zhao2023gecam}, the GECAM-C GRDs are placed on the top dome and bottom dome with different pointing angles (see Fig.\ref{SY-01_sat} and Fig.\ref{dome}). In fact, measurements of spectrum, timing and location of gamma-ray transients demand a detailed and accurate knowledge of the detector energy response. The GECAM-C GRDs energy response matrix is primarily derived from  Monte Carlo simulations based on the mass model of satellite and detectors, which need to be validated by calibration measurements.

In order to characterize the GECAM-C GRDs energy response, a series of on-ground calibration experiments have been carried out. The on-ground calibration results could also serve as the input for the in-flight calibration. All these calibrations and simulations are crucial for the energy response matrix. This paper focuses on the on-ground calibration campaigns of the GECAM-C GRDs, with emphasis on the analysis methods and results.

\begin{table}[H]
\centering
\caption{Main characteristics of GECAM-C GRDs.}
\begin{tabular}{ll} 
\hline \hline
Parameter & Value\\
\hline
Main detector & NaI(Tl); LaBr$_3$(Ce/Ce+Sr)\\
Detectors number & 12 \\
Energy range & 15-4000 keV \\
Detection area & 45.36 cm$^2$ \\
Energy resolution in on-ground test & < 18\% @59.5 keV \\
Deadtime & 4 $\mu s$ \\
Detection efficiency at 15 keV & > 75\% \\
\hline \hline
\label{GRD designed}
\end{tabular}
\end{table}

\section{Instrument description}
GECAM-C is a gamma-ray monitor developed with the payload technology and flight-spare hardware of GECAM-A and GECAM-B\cite{GECAMDesign1, GRDsDesign2}. All GECAM-C detectors are assembled on two dome modules which are installed on the top and bottom side of the satellite. Each dome module carries six Gamma-Ray Detectors (GRDs) and one Charged Particle Detector (CPD)\cite{CPDsDesign} (see Table.\ref{high-gain calibration}), with the former to detect gamma-rays and the latter to monitor charged particles. 

As the main detection unit, each GRD is composed of an encapsulated crystal box placed NaI(Tl) (NGRD for short) or LaBr$_3$(Ce/Ce+Sr) crystal (LGRD for short), Silicon Photomultipliers (SiPM), pre-amplification electric circuit and related mechanical structures. Both NaI(Tl) and LaBr$_3$(Ce/Ce+Sr) crystal have a diameter of 76 mm and thickness of 15 mm, manufactured by Beijing Glass Research Institute in China. They are fully encapsulated in aluminum box to maintain dry and avoid hygroscopy of crystal. The beryllium (Be) sheet with thickness of 200 $\mu m$ is used as the incident window, for the structural strength and the transmittance of low energy X-ray photons. The ESR film with thickness of 65 $\mu m$ and Teflon are applied as the reflective layer to wrap the top and the sides of crystal, respec         tively. The crystal is coupled with the quartz glass (thickness of 5 mm) by optical silicone. The quartz glass as light guild is also matched with the SiPM by silicone rubber optical interface. SiPM consists of many Avalanche Photo-Diode (APD) arrays working in Geiger mode to output a charge pulse signal after detecting photons.

\begin{figure}[H]
\centering
\includegraphics[width=15cm,height=5cm]{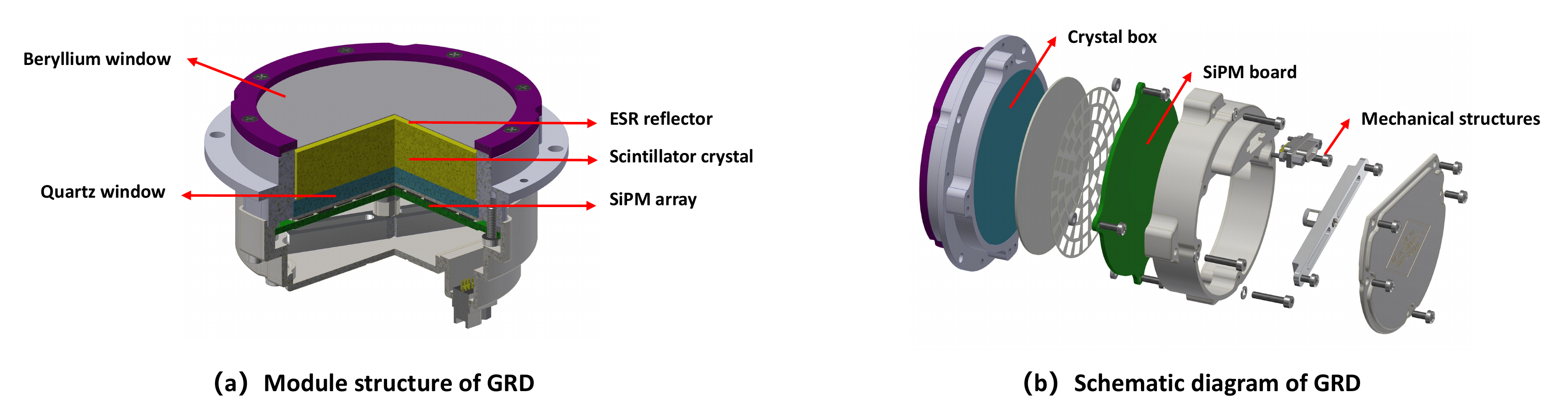}
\caption{GECAM-C GRD models.}
\label{GRDs}
\end{figure}

The SiPM array of each GRD makes use of 64 SiPM chips and each SiPM chip is 6.07×6.07 $mm$ in size but contains 22,292 micro-cells with pixel size of 35 $\mu m$. The pre-amplification electric circuit consists of a first-order preamplifier and a second-order preamplifier to achieve the measurement energy within a large dynamic range from keV to MeV\cite{Zhang2023}. Two types of electronic readout are used due to the limited data acquisition electronic resource: six LGRDs and four NGRDs utilize the dual cahnnels readout (high-gain channel and low-gain channel) while another two NGRDs are read out only by single channel (ZY-03 and ZY-07 detector)\cite{Zhang2023}. Fig.\ref{GRDs} shows the GRD structure.

\section{Calibration campaign}
The energy response matrix could be generated for each GRD with Geant4 simulation tool\cite{geant4}. 
To validate the simulation of energy response matrix, the actual detector response could be measured at discrete energies for some incident directions. 

The following subsections are dedicated to the description of the on-ground calibration campaigns on GECAM-C GRDs. The most complete calibration of all GECAM-C GRDs (high-gain channel) were calibrated with the Hard X-ray Calibration Facility (HXCF) %with the support and collaboration of 
at the National Institute of Metrology (NIM) \cite{HXCF_1,HXCF_2}. 
On the other hand, the low-gain channel calibration experiments were performed at Institute of High Energy Physics (IHEP) with a series of radioactive sources.

\subsection{Calibration with the X-ray beam}
\begin{figure}[H]
\centering
\includegraphics[width=16cm,height=8cm]{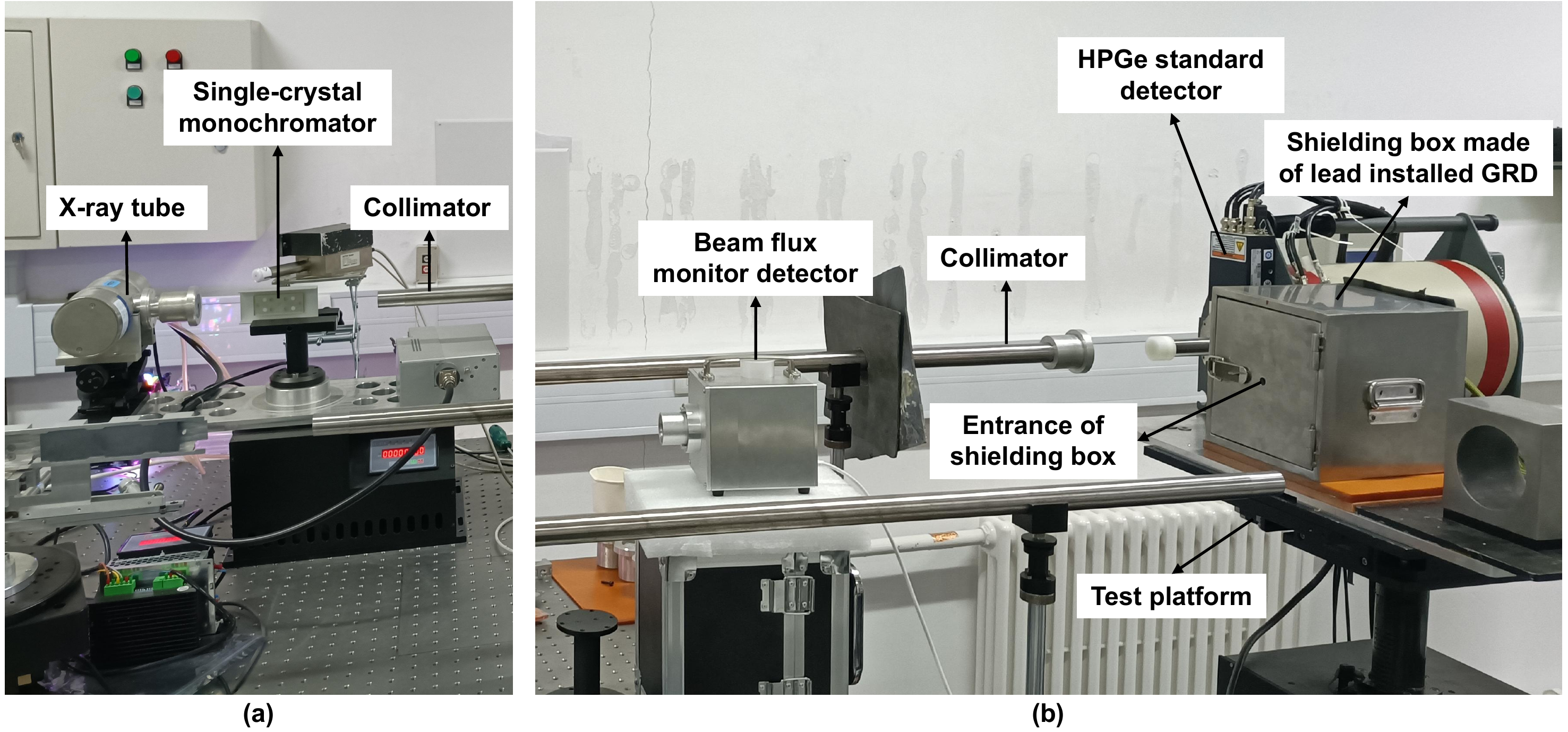}
\caption{Calibration experiment facility. HXCF is composed with an X-ray tube, crystal monochromators, collimators and X-ray beam monitor. Both the HPGe detector and the shielding box made of lead are fixed on the electric test platform.}
\label{HXCF}
\end{figure}

The HXCF is a calibration facility in the hard X-ray energy range which was firstly built for the High Energy telescope of HXMT\cite{2019Ground}. It has played an important role in the calibration of a series of gamma-ray detectors, such as GRID, GECAM-A and GECAM-B, SVOM/GRM etc \cite{GRID_calibration,gecam_calibration_GRD,SVOM_calibration_GRM}. The x-ray beam system consists of an X-ray tube, a single-crystal monochromator (5-40 keV), a double-crystal monochromator (20-140 keV), collimation system for shielding the stray light and limiting the size of the X-ray beam spot, an X-ray beam flux monitor detector to record the X-ray beam flux variation\cite{BeamMonitor}, an electric test platform, and a shielding box made of lead. 
Before the calibration campaign, a high-purity germanium (HPGe) detector should be calibrated accurately using a set of radioactive sources \cite{LEGe, qie2021study}. Then this HPGe detector is used to measure the spectrum and flux of the monochromatic X-ray beam. Detailed setup of the calibration experimental is shown in Fig.\ref{HXCF}.

A series of X-ray energies (see Table\ref{high-gain calibration}) were utilized to calibrate GRDs in high-gain channel. During the calibration test, the beam intensity, which depend on the voltage and current of X-ray tube, was set to less than 4000 cnts/s. GRD was installed in a lead shielding box and the HPGe detector was fixed by side, both of which could be moved by the electric platform. For each energy point, the X-ray beam test and background test are arranged for the same data collecting time. The environmental temperature of the laboratory was controlled at 22 ± 3$^{\circ}$C and the SiPM bias voltage was fixed on 26.5 V.  

\begin{table}[H]
\centering
\caption{GECAM-C GRDs installed information and calibration energy in high-gain channel.}
\begin{tabular}{lc|c|c|ll} 
\hline
\multicolumn{1}{|l|}{Dome Type} & \multicolumn{1}{|l|}{Location} & \multicolumn{1}{|l|}{Crystal ID} & \multicolumn{1}{|l|}{Detector type} & \multicolumn{2}{|l|}{Calibration X-ray Energy} \\
\hline
\multicolumn{1}{|c|}{\multirow{6}{*}{Top dome}} & 1-01L & ZY-81 & LGRD & \multicolumn{2}{|c|}{\multirow{12}{*}{\shortstack{LGRD: 9,10,12,13,14,15,\\19,22,25,28,31,34,37,38,\\39,40,45,50,60,70,80,90,\\100,120,140,356$(^{133}$Ba) keV \\ \\ \\ NGRD:10,12,14,16,19,21,\\25,28,30,32,33,34,36,38,\\40,45,50,60,70,80,90,100,\\120,140,356($^{133}$Ba) keV}}}\\
\cline{2-4}
\multicolumn{1}{|c|}{~} & 1-02N & ZY-01 & NGRD & \multicolumn{2}{|c|}{~} \\
\cline{2-4}
\multicolumn{1}{|c|}{~} & 1-03L & ZY-53 & LGRD & \multicolumn{2}{|c|}{~} \\
\cline{2-4}
\multicolumn{1}{|c|}{~} & 1-04N & ZY-02 & NGRD & \multicolumn{2}{|c|}{~} \\
\cline{2-4}
\multicolumn{1}{|c|}{~} & 1-05L & ZY-21 & LGRD & \multicolumn{2}{|c|}{~} \\
\cline{2-4}
\multicolumn{1}{|c|}{~} & 1-06N & ZY-03 & NGRD & \multicolumn{2}{|c|}{~} \\
\cline{1-4}
\multicolumn{1}{|c|}{\multirow{6}{*}{Bottom dome}} & 1-07L & ZY-87 & LGRD & \multicolumn{2}{|c|}{~} \\
\cline{2-4}
\multicolumn{1}{|c|}{~} & 1-08N & ZY-05 & LGRD & \multicolumn{2}{|c|}{~} \\
\cline{2-4}
\multicolumn{1}{|c|}{~} & 1-09L & ZY-16 & NGRD & \multicolumn{2}{|c|}{~} \\
\cline{2-4}
\multicolumn{1}{|c|}{~} & 1-10N & ZY-06 & LGRD & \multicolumn{2}{|c|}{~} \\
\cline{2-4}
\multicolumn{1}{|c|}{~} & 1-11L & ZY-17 & NGRD & \multicolumn{2}{|c|}{~} \\
\cline{2-4}
\multicolumn{1}{|c|}{~} & 1-12N & ZY-07 & LGRD & \multicolumn{2}{|c|}{~} \\
\hline 
\label{high-gain calibration}
\end{tabular}
\end{table}

\begin{figure}[H]
\centering
\includegraphics[width=6cm,height=6cm]{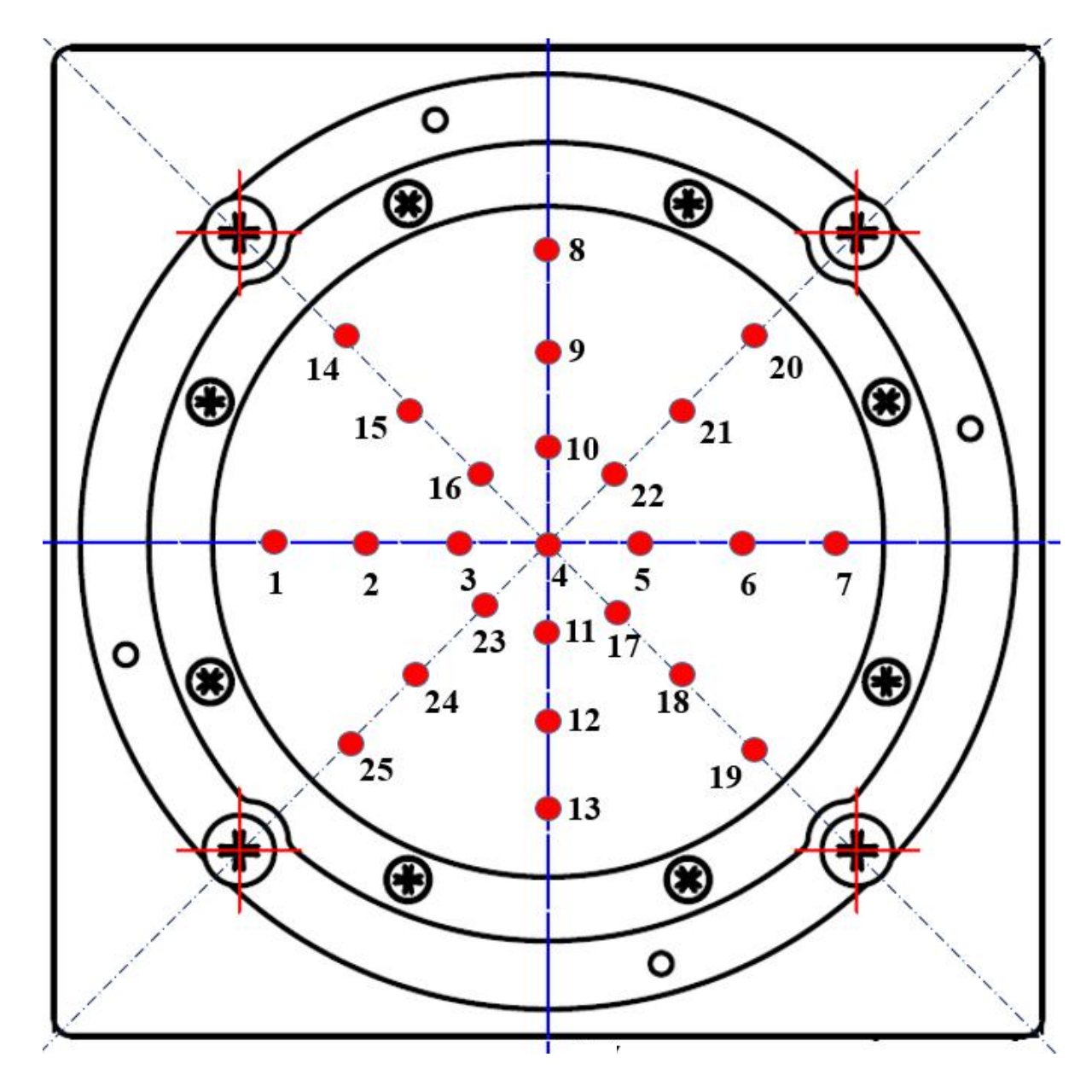}
\caption{The experimental setup of spatial non-uniformity test that shows the distribution of the scanning positions on the GRD surface.}
\label{spatial non-uniformity test}
\end{figure}

Furthermore, because of the non-uniformity of scintillation photon collection efficiency of the crystal and the sensitivity of the SiPM, energy response and full energy detection efficiency could be discrepant at different positions on the surface of the detector. To derive the spatial homogeneity of GRDs, we make use of a narrow X-ray beam (3-5 mm) to scan a set of discrete and uniform distributed points on the surface of the detector (see Fig.\ref{spatial non-uniformity test}). During the scanning process, each position was scanned by the X-ray beam, which recorded the X-ray beam spectrum and background spectrum successively.

\subsection{Calibration with the radioactive sources}
\begin{figure}[H]
\centering
\includegraphics[width=12cm,height=5cm]{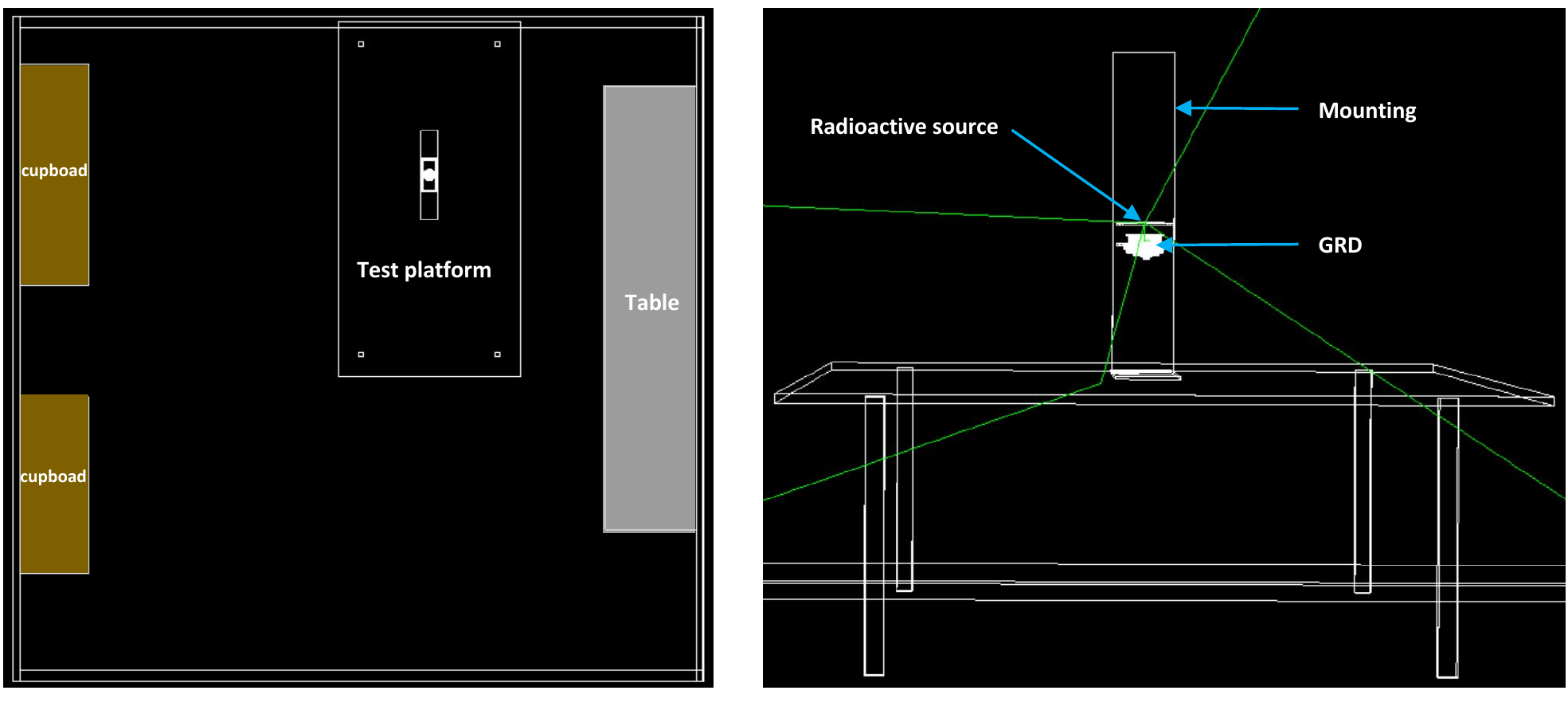}
\caption{The left panel shows in a top view of the laboratory main objects. The right panel shows the test platform in detail. A GRD is installed on the mounting and the radioactive source is placed above the GRD with distance 5 cm.}
\label{radioactive_source_model}
\end{figure}

The low-gain channel of GRDs calibration campaign was performed at the IHEP laboratory using a set of radioactive sources as listed in Table.\ref{radioactive source info} (see Fig.\ref{radioactive_source_model}). In addition, the SiPM bias-voltage and gain relation was also tested using $^{241}$Am and $^{137}$Cs radioactive sources. The SiPM bias-voltage varies from 25.6 V to 27.8 V with step of 0.2 V. During the radioactive source test, the radioactive source was placed at 0.05 m to the GRD. Besides, the intrinsic and environmental background was also recorded shortly before or after the measurements with radioactive sources. The laboratory temperature was controlled at 23 ± 2$^{\circ}$C.

\begin{table}[H]
\centering
\caption{Properties of radioactive sources used for GRDs calibration.}
\begin{tabular}{llllll} 
\hline \hline
Source & Half-life & Energy (keV) & Intensity & Activity (Bq) \\
\hline
$^{241}$Am & 432.6y (6) & 59.54 (1) & 35.9\% (4) & 9E$^3$  \\
$^{57}$Co & 271.74d (6) & 122.06 (12) & 85.60\% (17) & 1E$^4$  \\
$^{133}$Ba & 10.511y (11) & 356.01 (7) & 62.05\% (4) & 1E$^5$  \\
$^{22}$Na & 2.6018y (22) & 511.00 & 180.76\% (4) & 1E$^4$  \\
$^{137}$Cs & 30.08y (9) & 661.66 (3) & 85.10\% (20) & 9E$^3$  \\
$^{60}$Co & 1925.28d (14) & \makecell[l]{1173.23 (3) \\ \hline 1332.49 (4)} & \makecell[l]{99.85\% (3) \\ \hline 99.9826\% (6)} & 8E$^3$ \\
\hline \hline
\label{radioactive source info}
\end{tabular}
\end{table}

\section{Data analysis and results}
\subsection{Energy spectrum and fitting}
\begin{figure}[H]
\centering
\includegraphics[width=16cm,height=6cm]{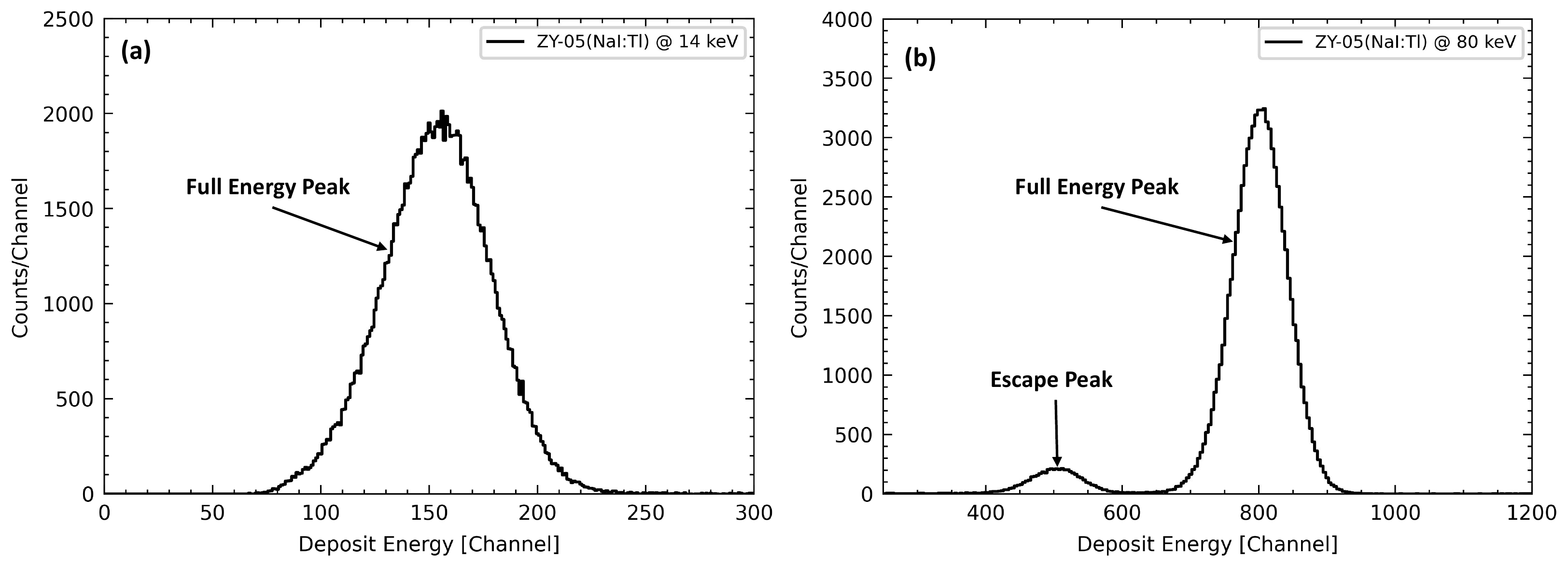}
\caption{Net spectrum with X-ray beam energy 14.0 keV and 80.0 keV for ZY-05 (NaI:Tl) detector in high-gain channel.}
\label{ZY-05_netspec_HXCF}
\end{figure}

\begin{figure}[H]
\centering
\includegraphics[width=16cm,height=6cm]{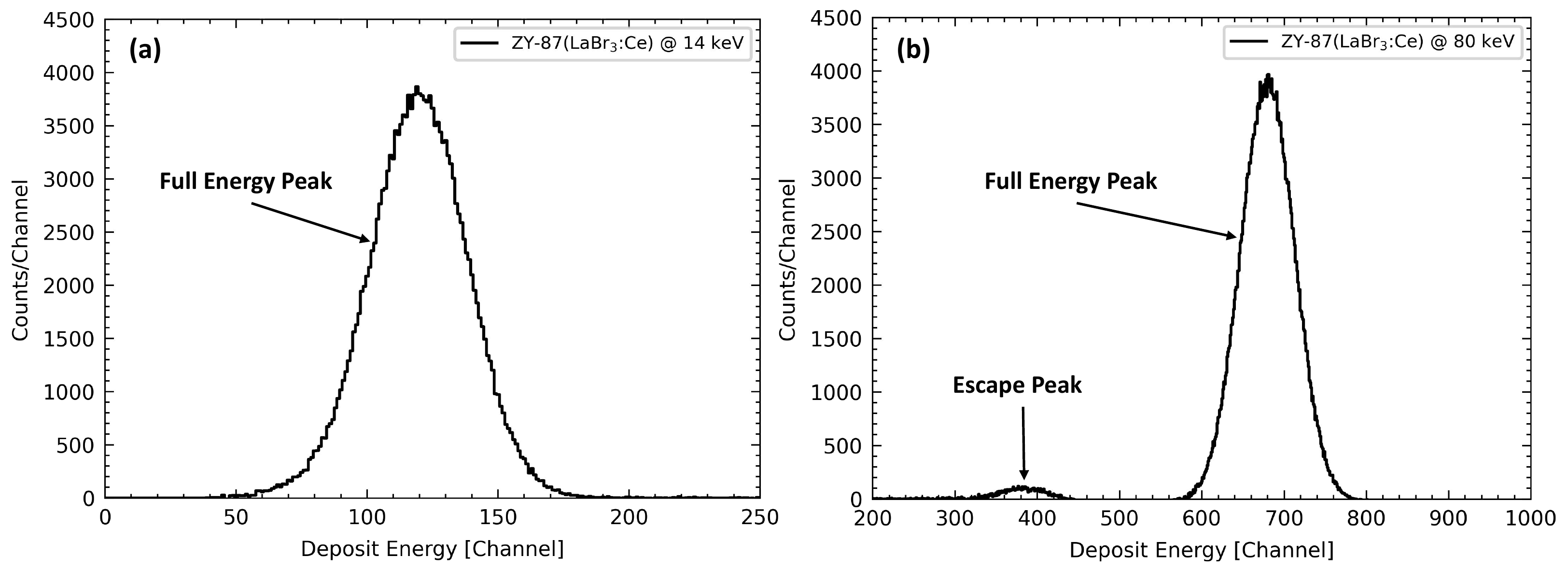}
\caption{Net spectrum with X-ray beam energy 14.0 keV and 80.0 keV for ZY-87 (LaBr$_3$:Ce) detector in high-gain channel.}
\label{ZY-87_netspec_HXCF}
\end{figure}

\begin{figure}[H]
\centering
\includegraphics[width=16cm,height=5cm]{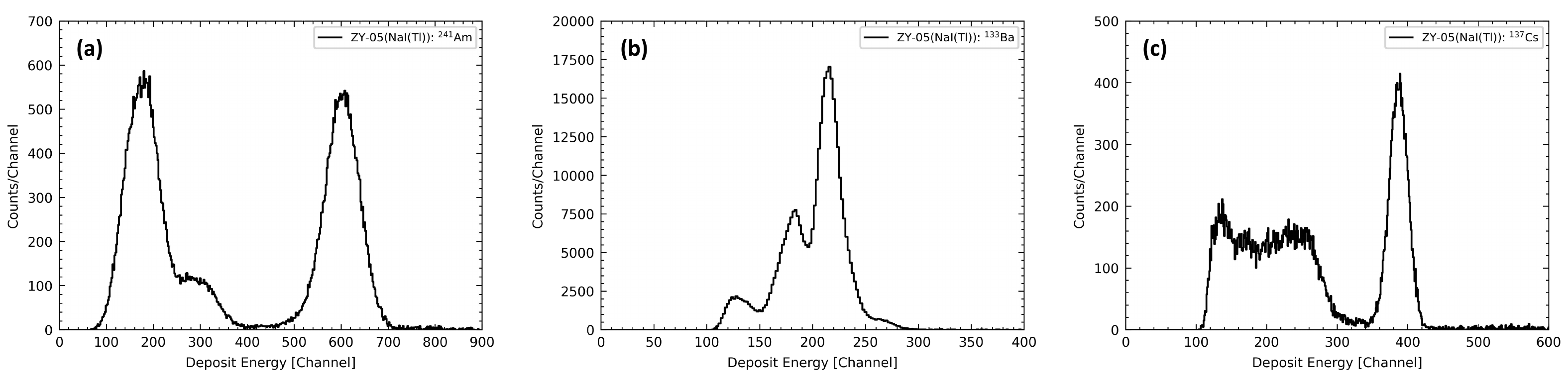}
\caption{Net spectrum for ZY-05 (NaI:Tl) detector measured with radioactive sources. (a) is the spectrum of $^{241}$Am(high-gain channel), (b) is the spectrum of $^{133}$Ba(low-gain channel), (c) is the spectrum of $^{137}$Cs(low-gain channel).}
\label{ZY-05_netspec_RS}
\end{figure}

\begin{figure}[H]
\centering
\includegraphics[width=16cm,height=5cm]{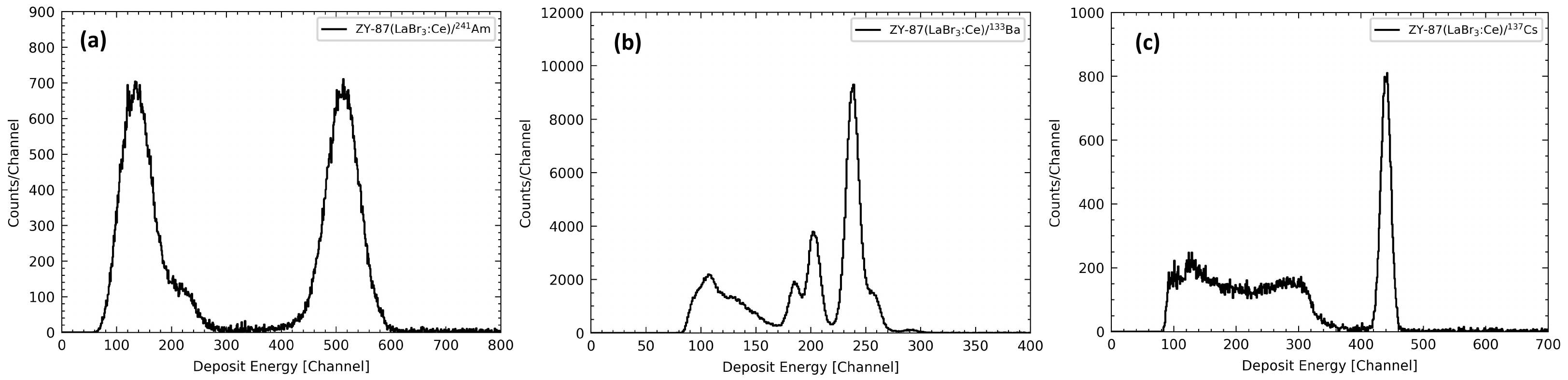}
\caption{Net spectrum for ZY-87 (LaBr$_3$:Ce) detector measured with radioactive sources. (a) is the spectrum of $^{241}$Am (high-gain channel), (b) is the spectrum of $^{133}$Ba (low-gain channel), (c) is the spectrum of $^{137}$Cs (low-gain channel).}
\label{ZY-87_netspec_RS}
\end{figure}

The net spectrum, subtracting background from the X-ray beam spectrum or radioactive source spectrum after the dead time correction, are shown from Fig.\ref{ZY-05_netspec_HXCF} to Fig.\ref{ZY-87_netspec_RS}, which highlight a distinctive feature of the LGRD and NGRD. Below the Bromine (Br) K-edge energy of 13.47 keV, only the full energy peak displayed. When the X-ray photons energy is higher than K-edge energy and also the escape peak deposited energy has exceeded the threshold, the escape peak of Br would appear on the left side of the full energy peak, as well as for Lanthanum (La) escape peak and Iodine (I) escape peak, for which the K-edge energy is 38.93 keV and 33.17 keV separately. The net spectrum could be fitted well by Gaussian function for the full energy peak and escape peak (Equ.\ref{Gaussians})\cite{Bhat2008Ground}, as shown in Fig.\ref{ZY-87_netspec_RS}. The reduced chi-square is applied to represent the goodness of fit, and the values of full energy peak in each spectrum are closing to 1. The fitting results provide important information for the energy response of the detector, such as the centroid of energy peak \textbf{x$_{c,i}$} ($Ch$), the standard deviation \textbf{$\sigma$}, the full width at half maximum \textbf{$w_i$} = 2.355 $\cdot$ $\sigma$, and the fitting errors etc. According to these fitting parameters, the energy resolution can be calculated as $Res$ = $\frac{2.355\cdot\sigma}{Ch}$ $\times$ 100 \% and the counts of full energy peak is computed only using the counts extracted within $Ch$ ± 2.58$\cdot$ $\sigma$ from the net spectrum. 

\begin{equation}
\begin{aligned} 
G(x;x_{c,i},w_i)=\sum \frac{A_i}{w_i} \cdot \sqrt{\frac{4 \ln 2}{\pi}} e^{{-4 \ln 2} \frac{\left(x-x_{c,i}\right)^{2}}{w_i^{2}}} (i=1,2,3...)
\label{Gaussians}
\end{aligned}
\end{equation}

\begin{figure}[H] 
\centering
\includegraphics[width=16cm,height=12cm]{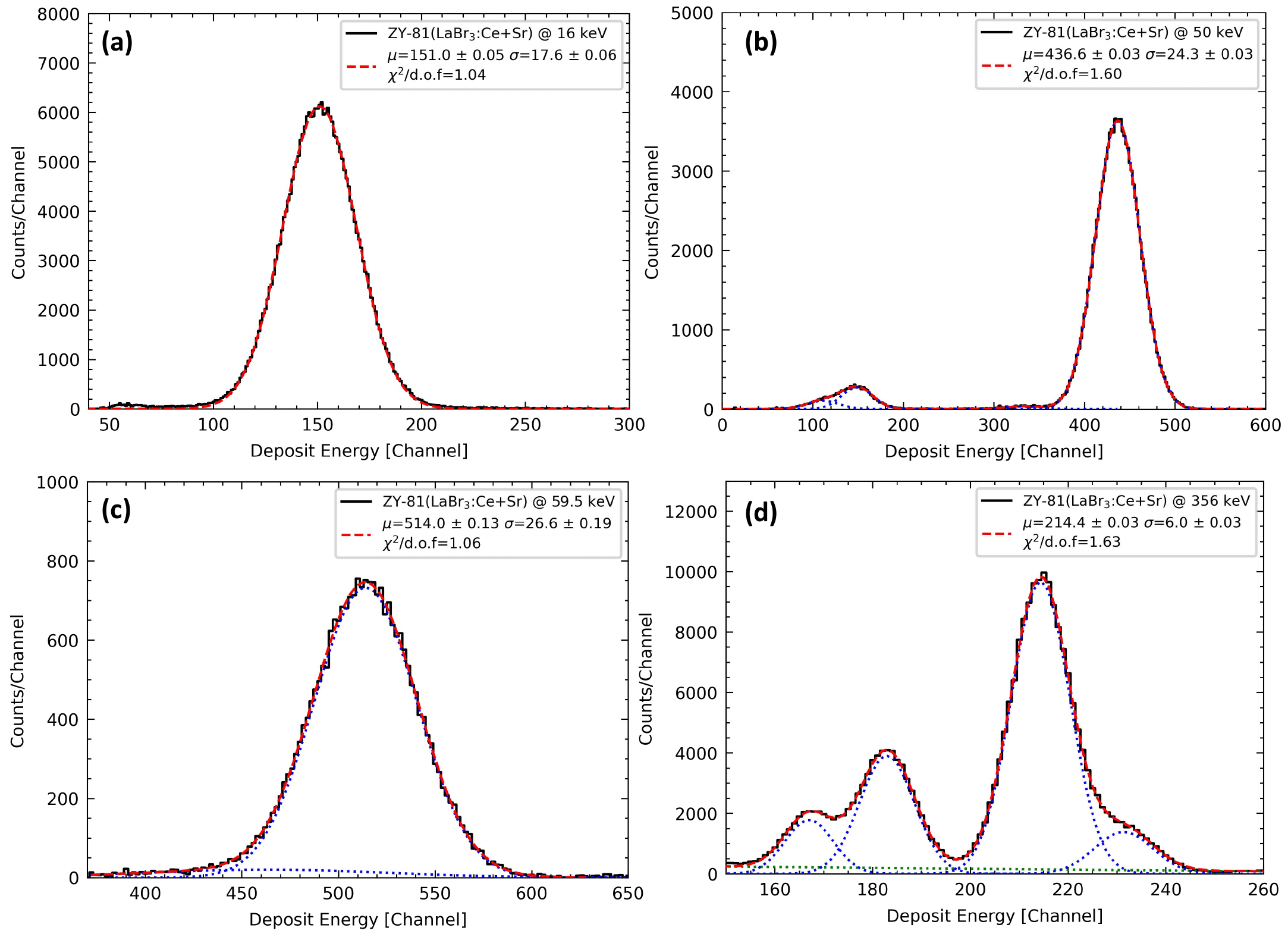}
\caption{The partial fitting results of net spectrum for ZY-81 (LaBr$_3$: Ce+Sr) detector. (a) and (b) is X-ray beam spectrum in high-gain channel. (c) is $^{241}$Am spectrum in high-gain channel. (d) is $^{133}$Ba spectrum in low-gain channel. The $\sigma$ and $\mu$ in four subplots represent full energy peak parameters with the photons energy writing in the label separately.}
\label{ZY-81_fitnetspec}
\end{figure}

The radioactive sources spectra are shown in Fig.\ref{ZY-05_netspec_RS} and Fig.\ref{ZY-87_netspec_RS}, which is more complicated with multi-components caused by Compton scattering in the detector and the radioactive source decay through other decay channels ($\alpha$, $\beta$ decay channel etc). Therefore, to fit radioactive source spectrum, some other functions (linear, quadratic or exponential function) should been added (besides the Gaussian function) to account for non-photo-peak contributions. Fig.\ref{ZY-81_fitnetspec} shows partial fitting results of ZY-81 (LGRD) detector spectrum measured by the HXCF and the radioactive sources.

\subsection{Channel–Energy relation}

\begin{figure}[H]
\centering
\includegraphics[width=12cm, height=8cm]{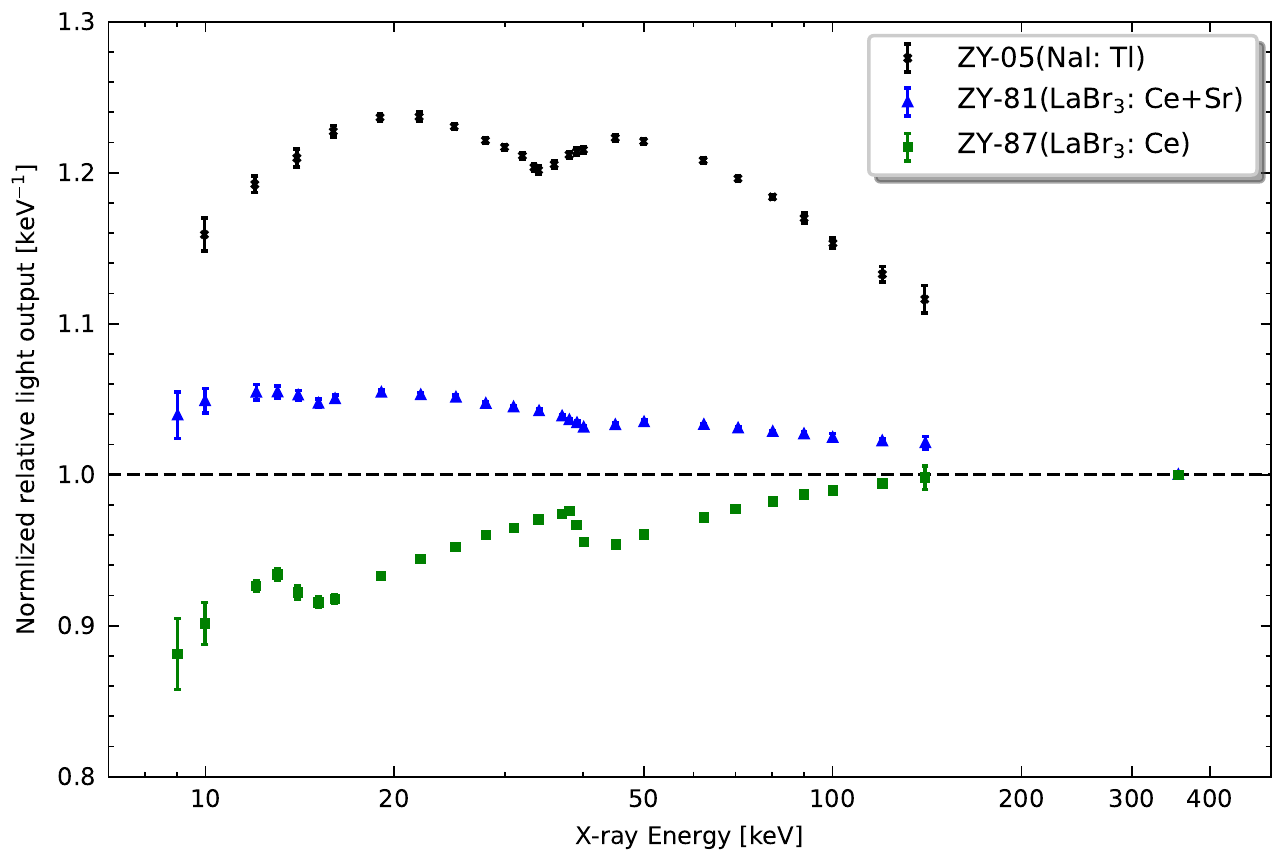}
\caption{The photon nPR of three flight GRDs as a function of X-ray energy, normalized to 356 keV.}
\label{nPR}
\end{figure}

Many experimental studies have indicated that the energy response of X/\textbf{$\gamma$}-ray photons both in NaI(Tl) and LaBr$_3$(Ce/Ce+Sr) crystal is non-proportional, because the scintillation efficiency varies with the incident energy.\cite{Scintillator_nPR_1, Scintillator_nPR_2, Scintillator_nPR_3}. Consequently, the non-proportional response ($nPR$) must be considered when relating the deposited energy (ADC channel) to the incident X/\textbf{$\gamma$}-ray photon energies (keV). The $nPR$ at X-ray photon energy $E_X$ in LGRD and NGRD was defined as the light output (equals to the centroid of full energy peak) divided by X-ray photon energy. Fig.\ref{nPR} shows the $nPR$ of three flight GRDs as a function of X-ray photons energy, which normalized to 356 keV. The data points include X-ray beam measurements at NIM together with additional radioactive source ($^{133}Ba$). The $nPR$ shows that three clear dips in plot at the K-shell binding energy of I (@ 33.17 keV), Br (@13.47 keV) and La (@38.93 keV) separately, which are determined by the energy distribution of deposited electrons generated by X-ray photons and corresponding electron light yield.\cite{Li2020, Zheng2022}. 

\begin{figure}[H]
\centering
\includegraphics[width=8cm, height=6cm]{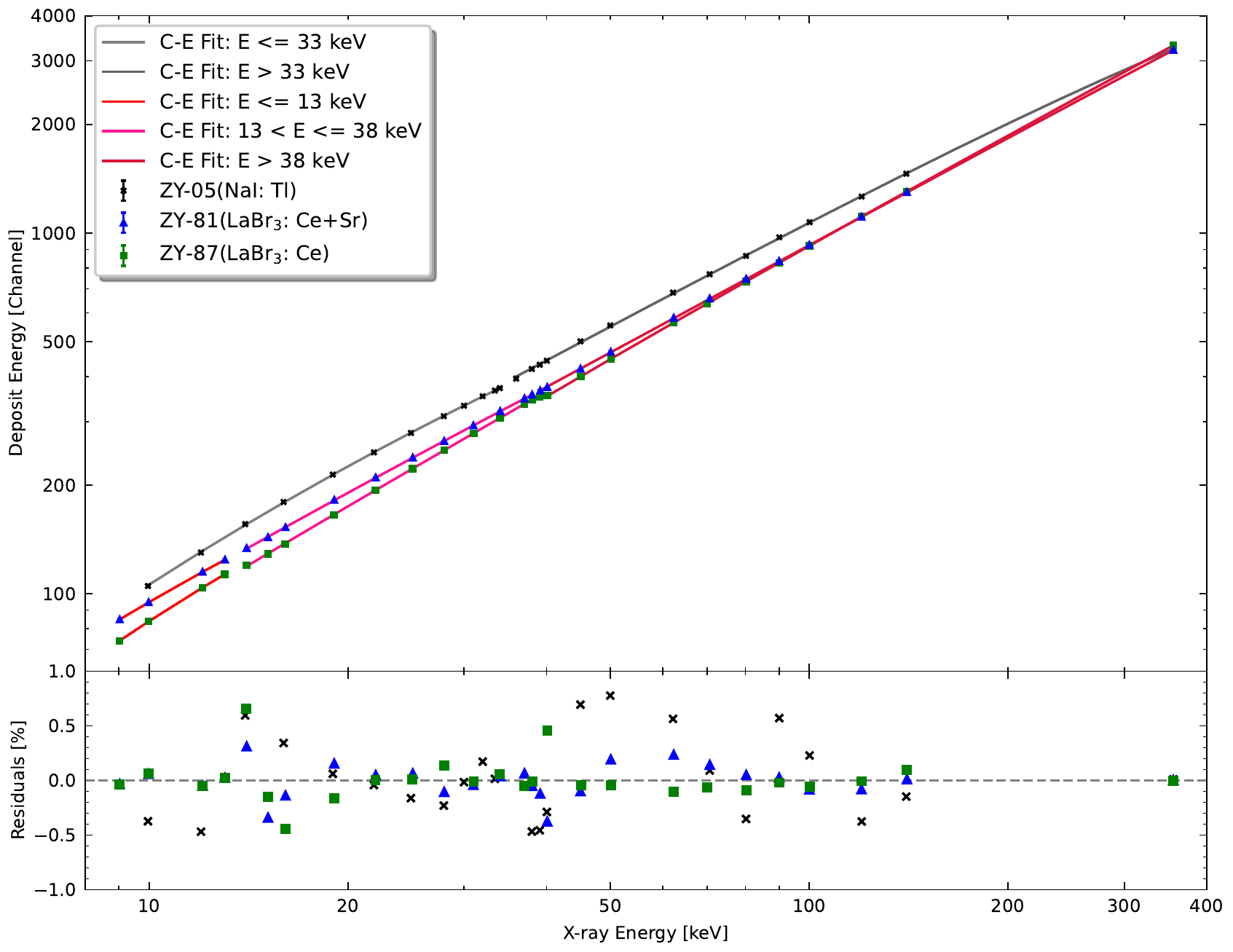}
\includegraphics[width=8cm, height=6cm]{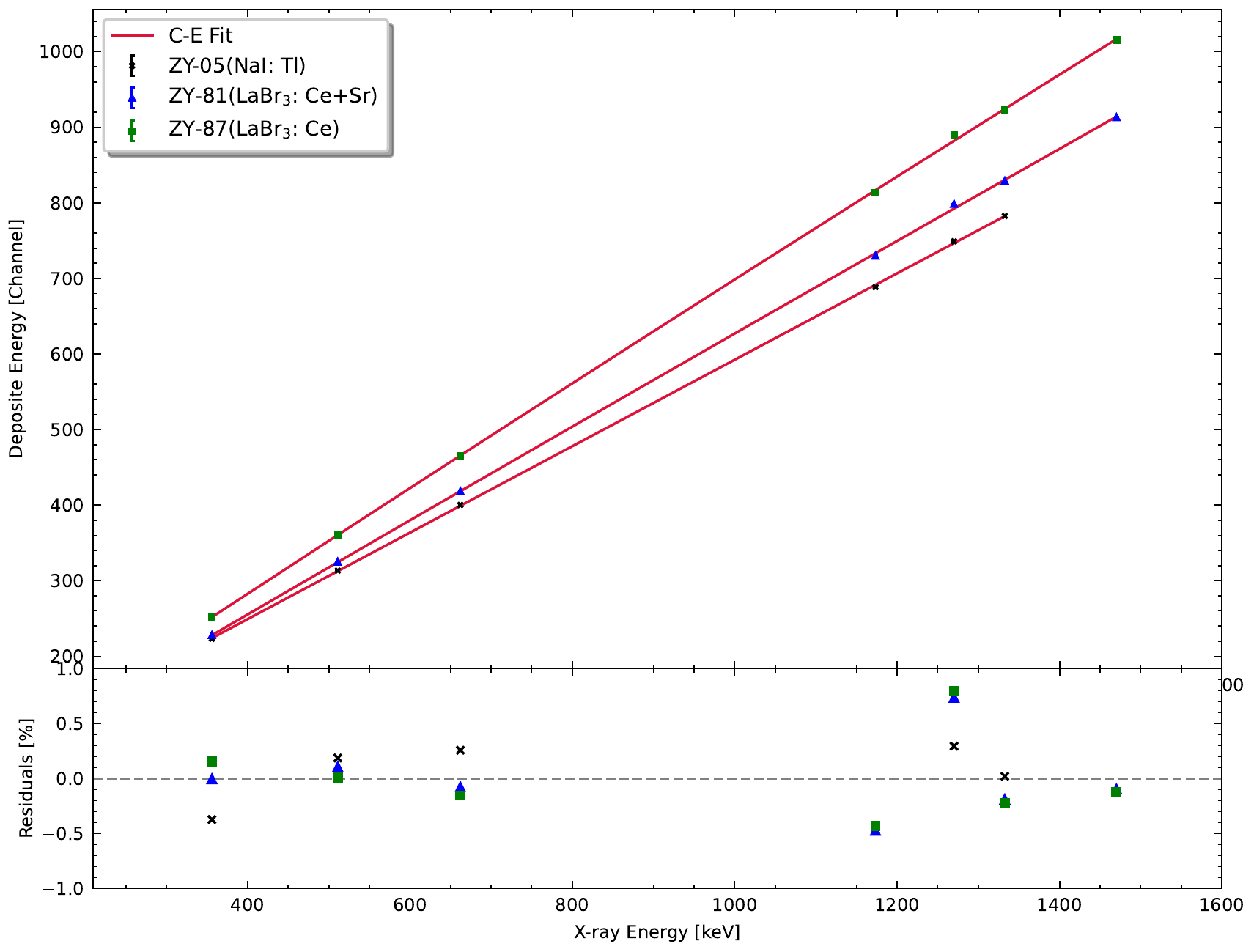}
\caption{Left panel: The Channel-Energy relation and fitting results of ZY-05 (NaT: Tl), ZY-81 (LaBr$_3$: Ce+Sr) and ZY-87 (LaBr$_3$: Ce) in high-gain channel separately. Right panel: The channel-Energy relation and fitting results of ZY-05 (NaT: Tl), ZY-81 (LaBr$_3$: Ce+Sr) and ZY-87 (LaBr$_3$: Ce) in low-gain channel separately.}
\label{EC_relation}
\end{figure}

According to the photons $nPR$ results, an integrated Channel-Energy relation could be fitted by Eq.\ref{quadratic} at different energy intervals which was separated by K-shell binding energy. The Channel-Energy relations of high-gain channel and low-gain channel were obtained as shown in Fig.\ref{EC_relation}. After considering the detection threshold and dynamic baseline, the high-gain channel was corresponds to the low range of detectable energy with ADC channels from about 50 to about 3400, and the low-gain channel corresponds to the high range of detectable energy with ADC channels from about 120 to about 3500, which cover the energy range from about 6 keV to 6 MeV according to the Channel-Energy relation.

\begin{equation}
\begin{aligned} 
Ch(E_\gamma) = b_2E_\gamma^2 + b_1E_\gamma + b_0
\label{quadratic}
\end{aligned}
\end{equation}

\subsection{Energy resolution}
\begin{figure}[H]
\centering
\includegraphics[width=8cm, height=6cm]{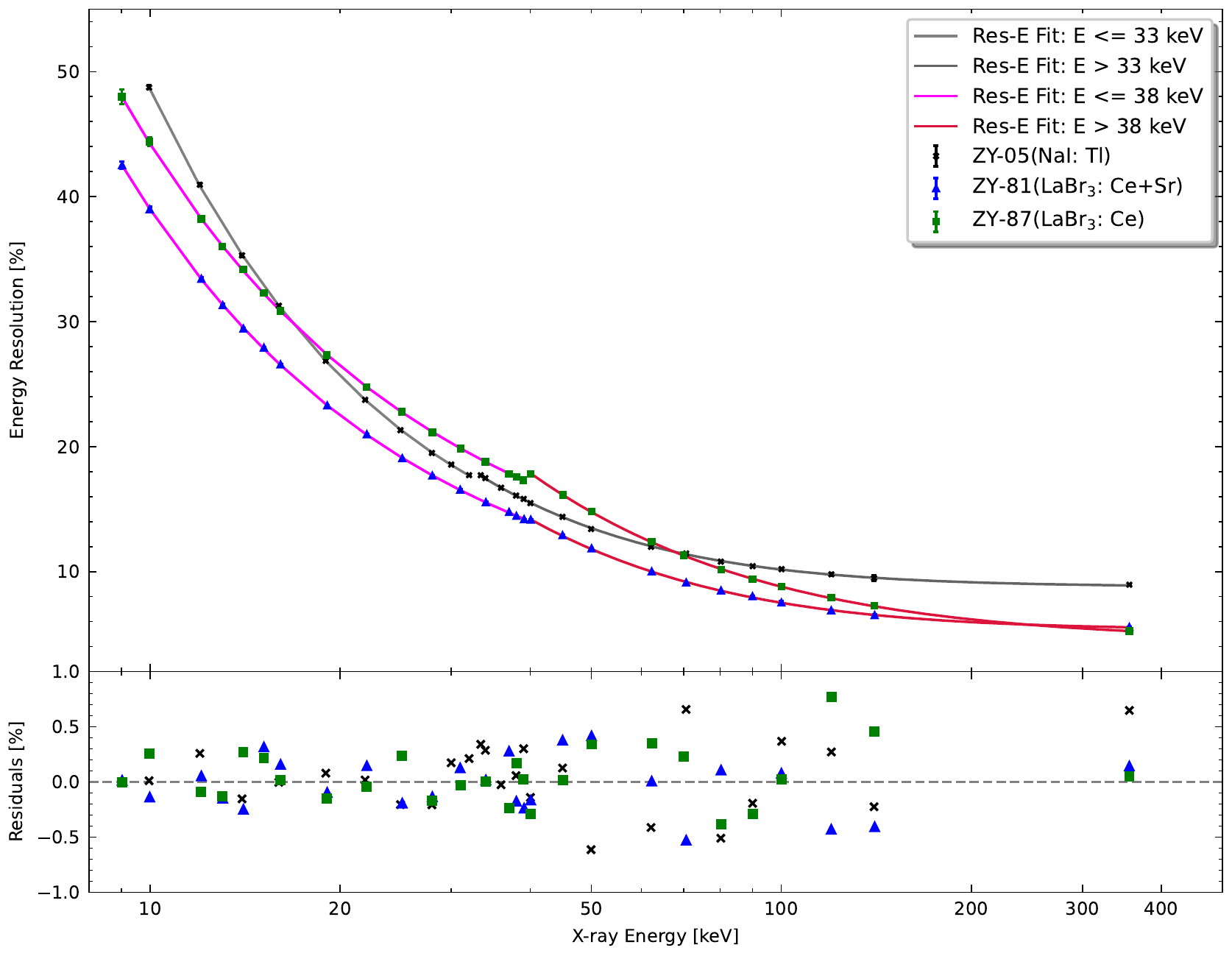}
\includegraphics[width=8cm, height=6cm]{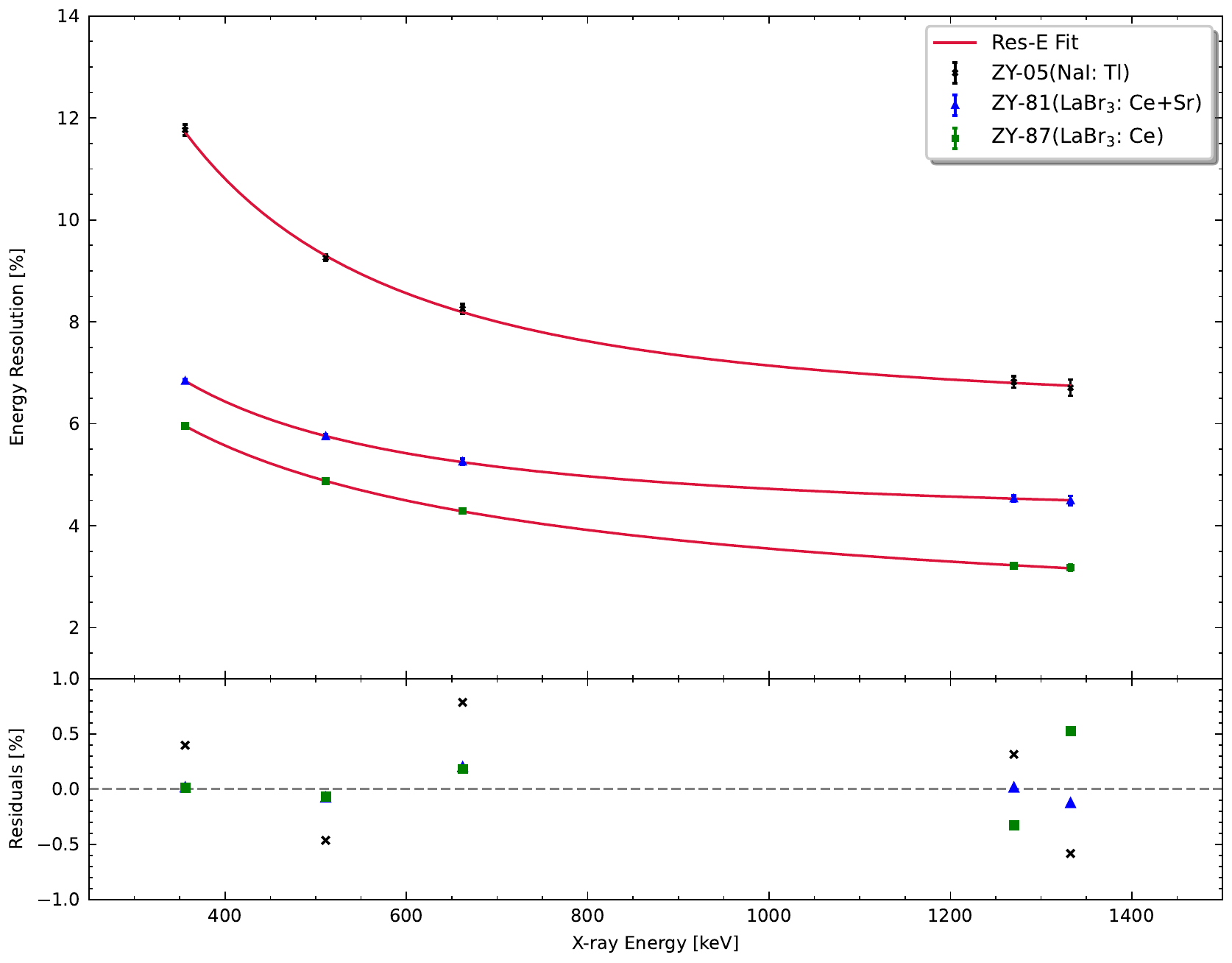}
\caption{Left panel: The intrinsic energy resolution at different X-ray photon energies and fitted by Equ.\ref{Res} for ZY-05 (NaI: Tl), ZY-81 (LaBr$_3$: Ce) and ZY-87 (LaBr$_3$: Ce+Sr) in high-gain channel respectively. Right panel: The intrinsic energy resolutions at different energies and fitted by Equ.\ref{Res} for ZY-05 (NaI: Tl), ZY-81 (LaBr$_3$: Ce) and ZY-87 (LaBr$_3$: Ce+Sr) in low-gain channel respectively.}
\label{Res_relation}
\end{figure}

\begin{figure}[H]
\centering
\includegraphics[width=12cm, height=6cm]{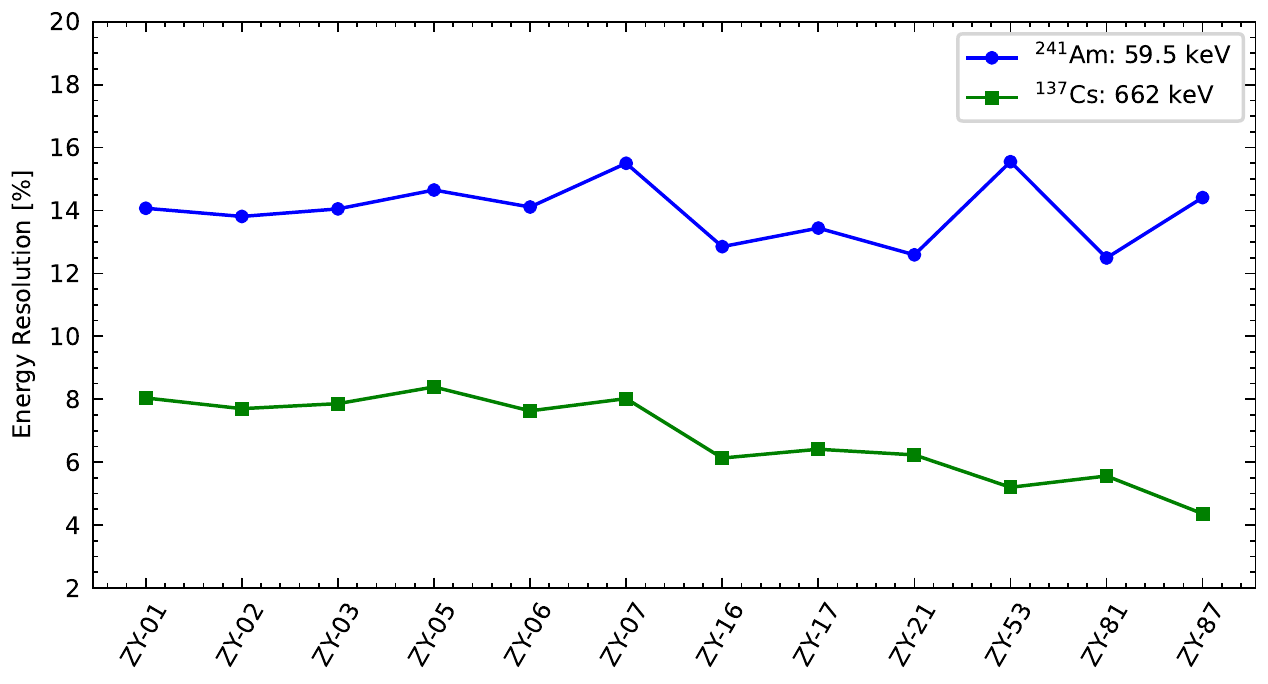}
\caption{The $^{241}$Am and $^{137}$Cs main peak energy resolution of each GRDs.}
\label{Am&CsRes}
\end{figure}

The intrinsic energy resolution of GRD can be inferred from the intrinsic $FWHM$ of full energy peak divided by its centroid of full energy peak shown as Eq.\ref{Res}\cite{SVOM_calibration_GRM}, where the constant term ($a$) reflects the electronic noise, the second term ($b$) is attributed to the statistical fluctuation of scintillating photons and photoelectrons, the third term ($c$) represents the non-proportionality of response ($nPR$) of scintillators\cite{Li2023, Fermi_GBM}. The intrinsic $FWHM$ of full energy peak equals to $2.355\cdot\sqrt{\sigma^2_{GRD} - \sigma^2_{beam}}$, where $\sigma_{GRD}$ is the standard deviation of the full energy peak obtained by fitting net spectrum and $\sigma_{beam}$ is the X-ray beam energy broaden tested by HPGe detector. The error of $Res$ was calculated by the error propagation function Eq.\ref{errs}, where the $x$ is the values of full energy peak centroid($Ch$), $y$ is the intrinsic standard deviation ($\sigma$) of GRD, corresponding to the errors of $\sigma_x$ and $\sigma_y$, and $\sigma$(u) is the expected error of resolution which is $u$. Fig.\ref{Res_relation} shows the intrinsic energy resolution at different X/$\gamma$-ray photon energies and the fitting results, which shows that all the residuals ((data-model)/data) are less than 1\%. Fig.\ref{Am&CsRes} indicate the $^{241}$Am and $^{137}$Cs main peak energy resolution of each GRDs and all GRDs energy resolution at 59.5 keV satisfy the design objective lower than 18\%.

\begin{equation}
Res(E_\gamma) = \frac{2.355\cdot\sigma(E_\gamma)}{Ch(E_\gamma)} =\frac{\sqrt{a^2+b^2E_\gamma+c^2E_\gamma^2}}{E_\gamma}\times100\%
\label{Res}
\end{equation}

\begin{equation}
\frac{\sigma^2(u)}{u^2}=\frac{\sigma_x^2}{x^2} + \frac{\sigma_y^2}{y^2}
\label{errs}
\end{equation}

\subsection{Detection efficiency}
\begin{figure}[H]
\centering
\includegraphics[width=16cm, height=6cm]{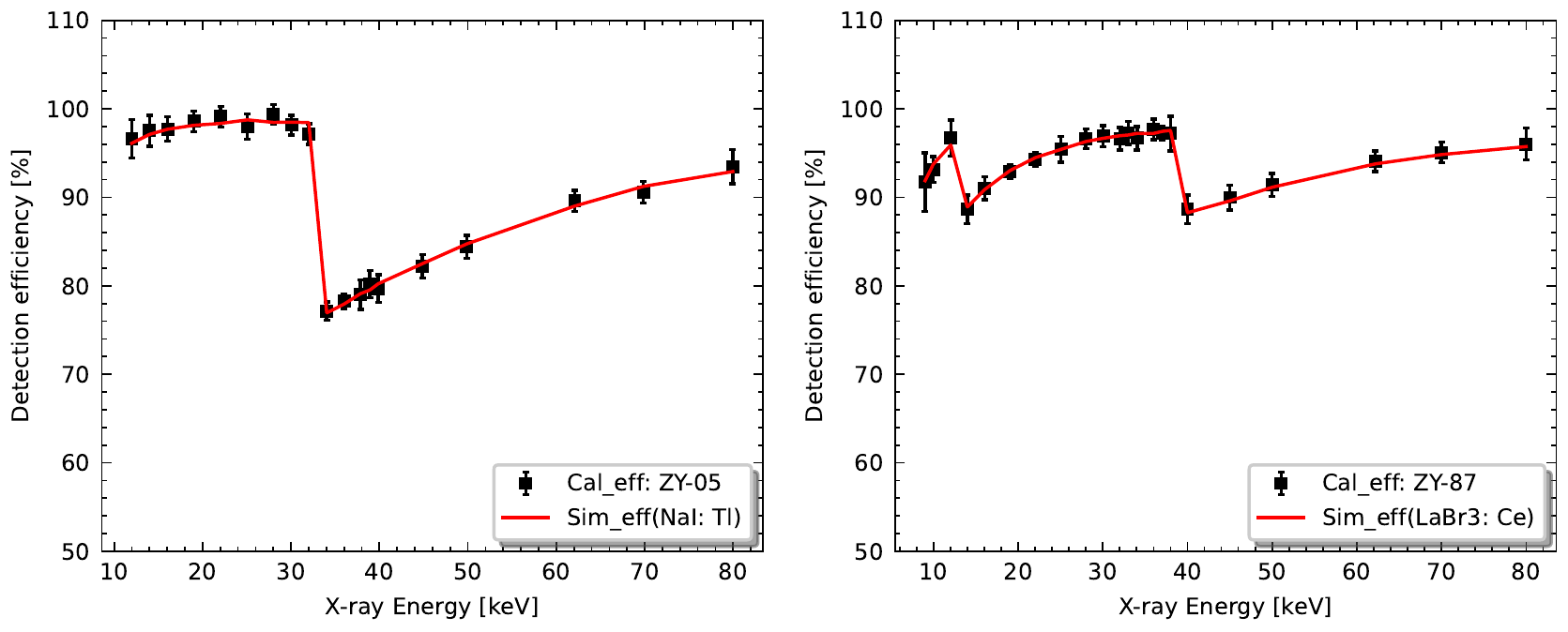}
\caption{The comparison intrinsic efficiency between experimental and Geant4 simulation at full energy peak.}
\label{Detection_efficiency}
\end{figure}

The intrinsic detection efficiency of full energy peak were determined by the calibration campaigns with HXCF and computed according to Equ.\ref{detection_efficiency_func}:

\begin{equation}
\epsilon = \frac{n_{GRD}(E_X)}{I(E_X)\cdot\kappa_I\cdot t} =
\frac{n_{GRD}(E_X)\cdot\epsilon_{HPGe}(E_X)}{n_{HPGe}(E_X)\cdot\kappa_I},
\label{detection_efficiency_func}
\end{equation}

where (1) $n_{GRD}(E_X)$ equals to the full energy peak area counts in 2.58$\cdot\sigma$ measured by GRD, (2)intrinsic X-ray beam flux $I(E_X)=\frac{n_{HPGe}(E_X)}{\epsilon_{HPGe}(E_X)\cdot t}$, where $n_{HPGe}(E_X)$ equals to the full energy peak area counts in 2.58$\cdot\sigma$ detected by the HPGe detector and $\epsilon_{HPGe}(E_X)$ is the detection efficiency of HPGe detector calibrated in advance\cite{LEGe}, (3) $\kappa_I$ represents the beam stability and was giving by the beam flux monitor detector in this work. The results of intrinsic detection efficiency of full energy peak as a function of the X-ray photons energy for NaI(Tl) and LaBr$_3$(Ce/Ce+Sr) GRDs are shown in Fig.\ref{Detection_efficiency}. Obviously, the measured detection efficiencies of GRDs is consistent with the simulation results derived from Geant4 within the margin of error.

\subsection{Relation of energy response and bias-voltage}
\begin{figure}[H]
\centering
\includegraphics[width=16cm, height=6cm]{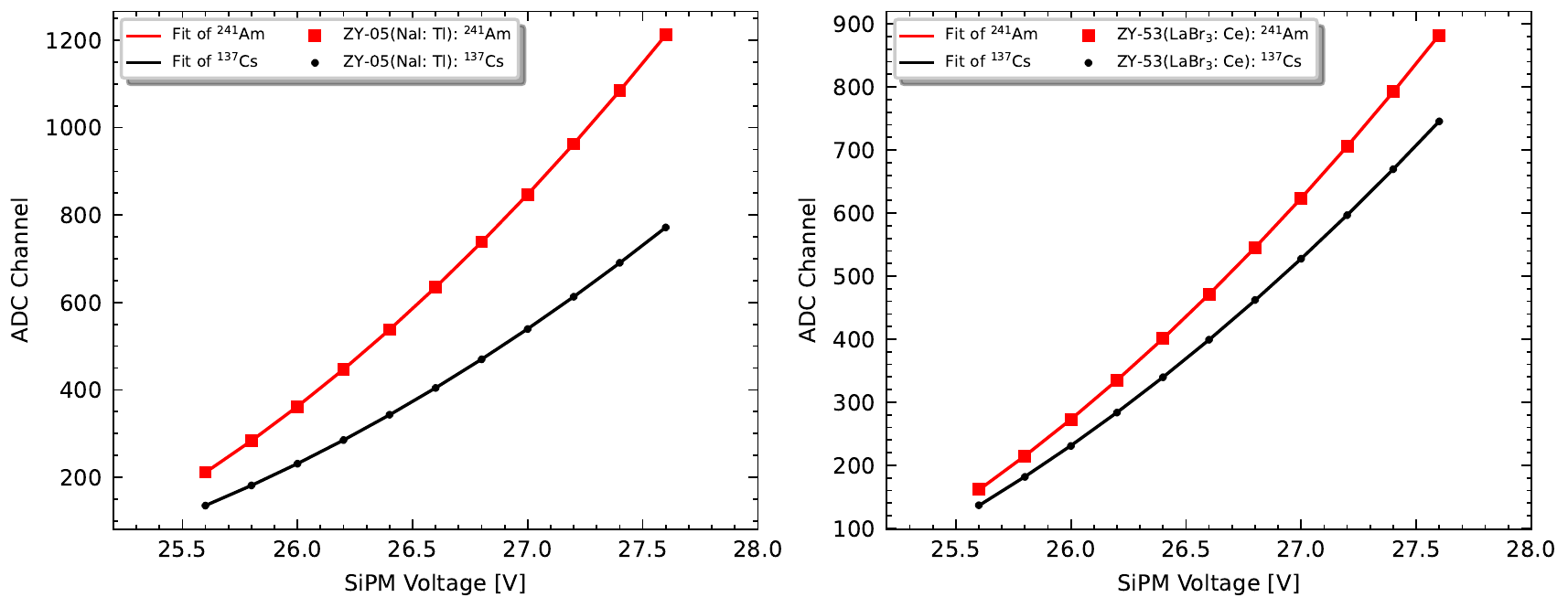}
\caption{The relation between full energy peak and SiPM bias voltage on NaI(Tl) detector (left panel) and LaBr$_3$(Ce) detector (right panel), which calibrated by $^{241}$Am at 59.5 keV (in high-gain channel), $^{137}$Cs at 662 keV (in low-gain channel).}
\label{Voltage_channel}
\end{figure}

\begin{figure}[H]
\centering
\includegraphics[width=16cm, height=6cm] {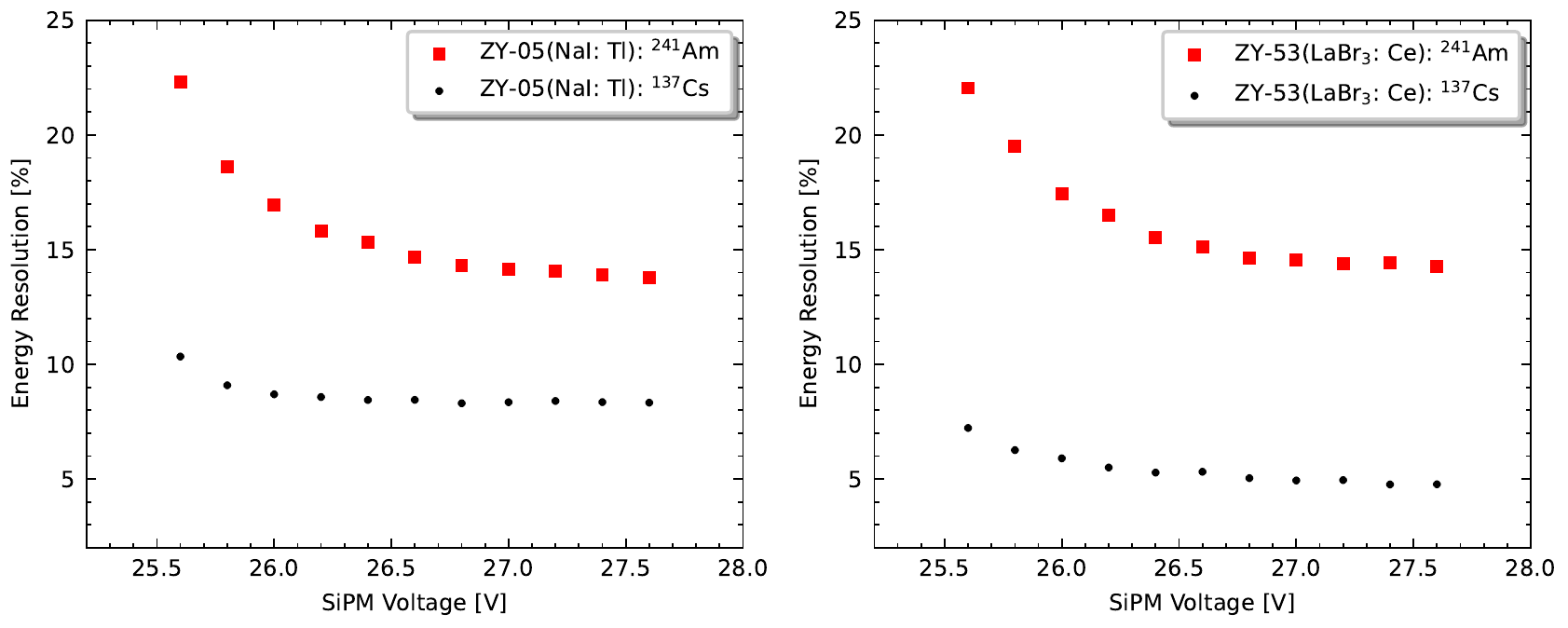}
\caption{The relation between energy resolution and SiPM bias voltage on NaI(Tl) detector (left panel) and LaBr$_3$(Ce) detector (right panel), which calibrated by $^{241}$Am at 59.5 keV (in high-gain channel), $^{137}$Cs at 662 keV(in low-gain channel).}
\label{Voltage_res}
\end{figure}

For different bias voltage of SiPM, the relation between full energy peak and SiPM bias voltage could be described well by a quadratic function. The full energy peak and energy resolution varying with bias voltage are shown in Fig.\ref{Voltage_channel} and Fig.\ref{Voltage_res}, which suggests that stabilizing gain drifting could be realized by adjusting SiPM bias voltage, and the energy resolution also could be improved by increasing SiPM bias voltage. Besides, the SiPM bias voltage and temperature response are also verified in detail to adjust gain drifting which is caused by temperature variation and irradiation damage on SiPM on-orbite\cite{Zhang2023}.

\subsection{Spatial non-uniformity results}
\begin{figure}[H]
\centering
\includegraphics[width=16cm, height=6cm]{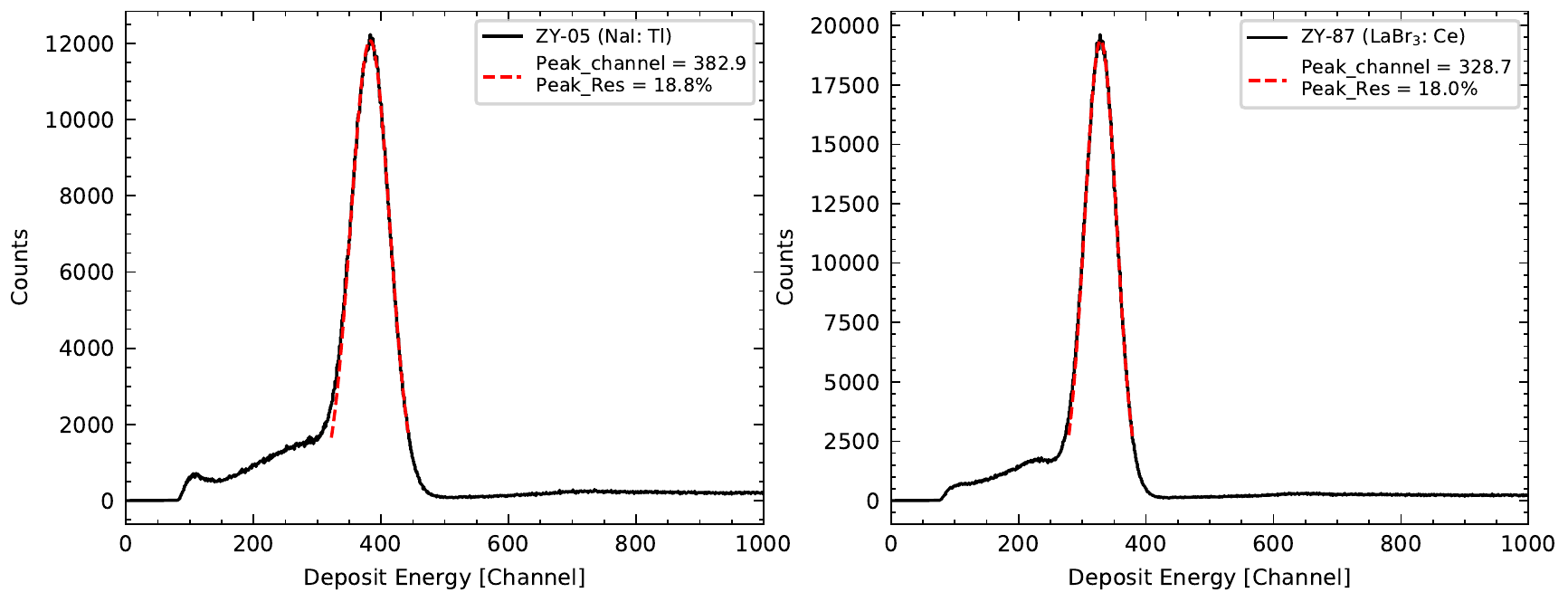}
\caption{Integrated spectrum for 25 scanning positions with X-ray energy 39 keV was used to describe the overall performance of GRDs. The left panel is ZY-05 (NaI: Tl) measurements and the right panel is ZY-87 (LaBr$_3$: Ce
) measurements.}
\label{Spatial_uniformity_spectra}
\end{figure}

\begin{figure}[H]
\centering
\includegraphics[width=16cm, height=18cm]{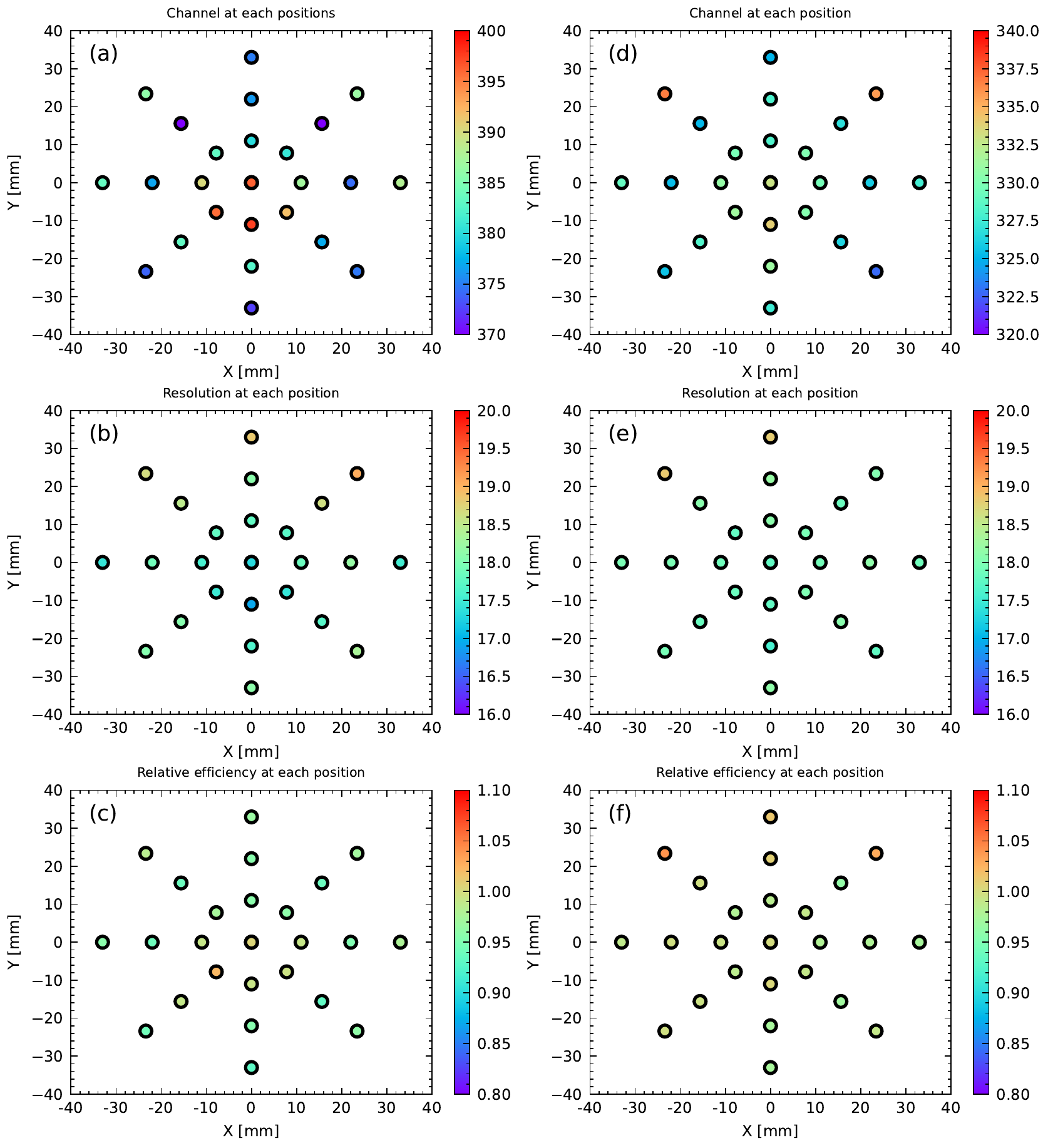}
\caption{The performance of each scanning positions. The figure (a), (b) and (c) is the peak channel, energy resolution and relative detection efficiency for ZY-05 (NaI: Tl) detector at energy 39 keV, respectively. The figure (d), (e) and (f) is the peak channel, energy resolution and relative detection efficiency for ZY-87 (LaBr$_3$: Ce) detector at energy 39 keV, respectively.}
\label{Spatial_uniformity}
\end{figure}

As mentioned in section 3.1, a total of 25 positions on the surface of GRDs (ZY-05 and ZY-53) have been scanned utilizing X-ray beam with energy 39 keV. After time normalization, the measured spectrum for each position and the integrated spectrum (shown in Fig.\ref{Spatial_uniformity_spectra}) was fit with Gaussian function to achieve the peak channel and energy resolution. Then, the centroid of full energy peak, energy resolution and relative detection efficiency (normalized to the value of central position) of each scanning positions (see Fig.\ref{Spatial_uniformity}) were computed separately. Besides, the spatial uniformity of detectors also could be evaluated using two parameters: (1) $\frac{(Max-Min)}{(Max+Min)}$, where the $Max$ and $Min$ was the maximum and minimum value of all the positions test results; (2) relative standard deviation (RSD), which equal to the standard deviation divided by the arithmetic mean of data, which reported in Table.\ref{SU}. As shown in Table.\ref{SU} and Fig.\ref{Spatial_uniformity}, the peak values, energy resolutions and relative detection efficiencies adequately proved that there was no obvious dead area of the crystal and the crystals were also not deliquesce on the edge.

\begin{table}[H]
\centering
\caption{The values representation the spatial uniformity of GRDs.}
\begin{tabular}{cccc} 
\hline \hline
  & Peak mean value & Energy resolution & Relative detection efficiency\\
\hline
$\frac{Max-Min}{Max+Min}$ of ZY-05 & 0.036 & 0.058 & 0.047 \\
RSD of ZY-05 & 0.022 & 0.030 & 0.026 \\
$\frac{Max-Min}{Max+2Min}$ of ZY-87 & 0.021 & 0.034 & 0.039 \\
RSD of ZY-87 & 0.010 & 0.016 & 0.017 \\
\hline \hline
\label{SU}
\end{tabular}
\end{table}

\section{Energy Response Validation: Simulation V.S. Calibration }
\begin{figure}[H]
\centering
\includegraphics[width=14cm, height=10cm]{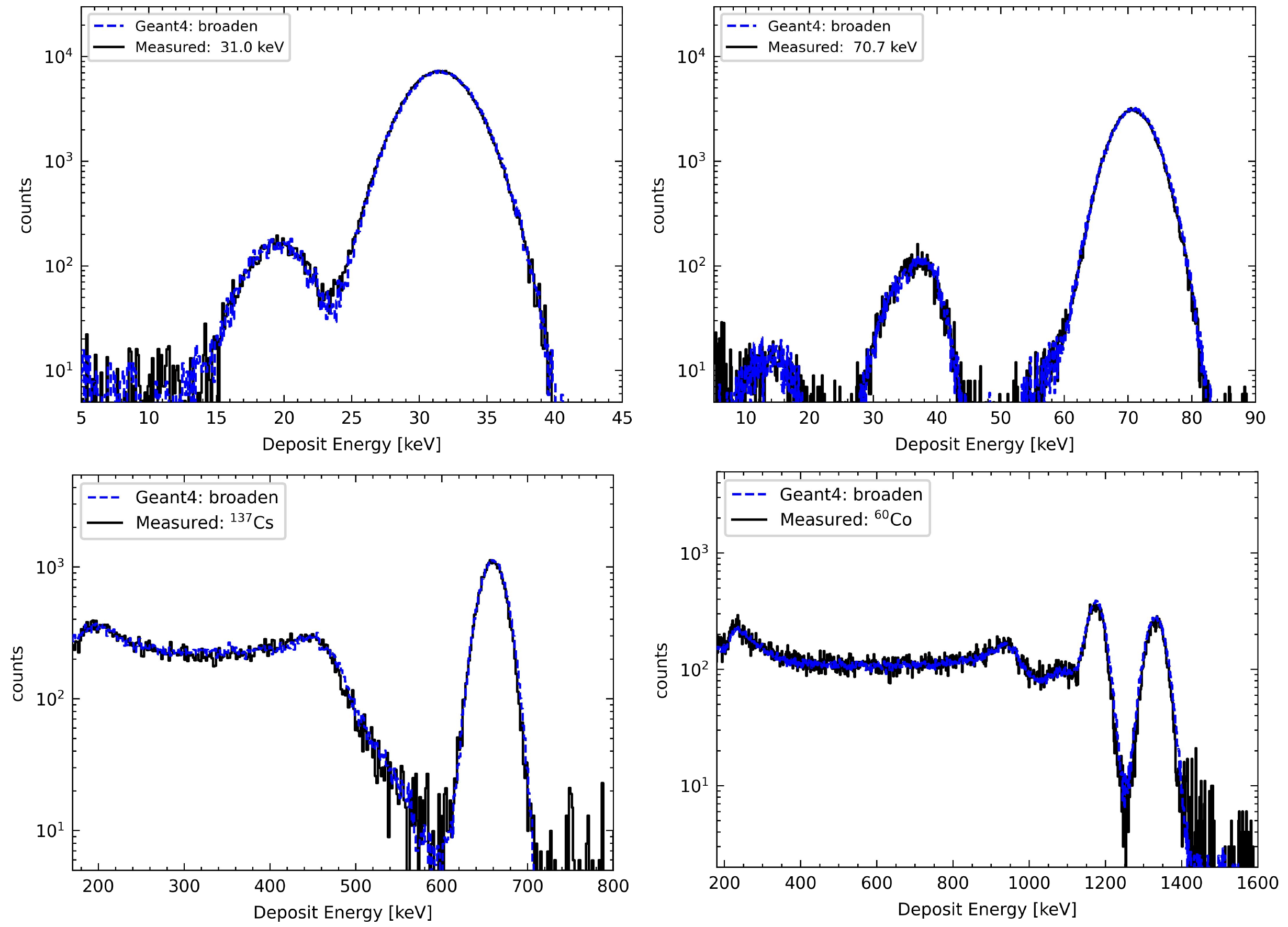}
\caption{The simulated spectrum compared with the measured spectrum under the irradiation of HXCF (31.0 keV \& 70.7 keV) and radioactive source ($^{137}$Cs \& $^{60}$Co) respectively.}
\label{SimVSCal}
\end{figure}

we developed a GEANT4-based code (version 4.10.6) to check the accuracy of the obtained channel-energy relation and energy resolution, as well as access accuracy of the GRD mass model. In order to realistically simulate all the radiation that reaches the detector, it was necessary to include all scattered radiation from the surrounding material near and far, which required a detailed modeling of the laboratory, such as GRD, brackets, desk, walls, cement floor, air, etc. The GRD mass model was converted directly from CAD and was imported into GEANT4. $G4EmLivermorePhysics$, a physical process of electromagnetic, is included in the physical list, and the $G4DecayPhysics$ and $G4RadioactiveDecayPhysics$ are enabled. In each run, the photons deposited energy in GRD sensitive detector were be recorded. For comparison purpose, the simulation spectrum broadened according to energy resolution was compared with measured spectrum which utilizing the Energy-Channel relation to convert channel to energy. Figure.\ref{SimVSCal} displayed partial results of the simulation spectrum (in blue and red) and measured spectrum (in black). It shows that the simulated spectra are good agreement with experimental spectra, including all the features such as the full-energy peaks, the escape peaks, the back-scattering peaks, the Compton edges, etc\cite{Zheng2022}. The non-parametric test method, Kolmogorov-Smirnov test, was implemented to test whether a significant difference between the two observed distributions, which utilized to verify the coincidence degree of simulation spectrum and measured spectrum. The python package ($scipy.stats.ks\_2samp$\footnote{\href{ https://docs.scipy.org/doc/scipy/reference/generated/scipy.stats.ks_2samp.html}{https://docs.scipy.org/doc/scipy/reference/generated/scipy.stats.ks\_2samp.html}}) was called to calculate the $P$ value. We got $P_1$=0.64 (HXCF @ 31.0 keV), $P_2$=0.57 (HXCF @ 70.7 keV), $P_3$=0.55 ($^{137}$Cs), $P_4$=0.45 ($^{60}$Co), which were more higher than significance level $\alpha$=0.05. It suggested that the simulation spectra and measured spectra obeying the same distribution. In other words, our energy response and GRD mass model are reliable.

\section{Conclusion}
In this work, we present the ground calibration of GECAM-C GRDs with X-ray beam and radioactive sources. The Channel-Energy relations, energy resolution relations, detection and spatial uniformity of the each GRD are measured in detail. The results show GRDs perform as expected. It is worth noting that the Channel-Energy relations can not be directly used for in-flight calibration database because of the different operating temperature. According to the in-flight energy calibration with characteristic lines in detectors, the GRD detection energy is from about 6 keV to 6 MeV\cite{Zhang2023}. All those measurements could be used in the validation of Monte Carlo simulations of the GECAM-C GRDs detector response. 

With the wide energy range, large field of view, moderate localization capability and real time trigger alerts,  GECAM-C is expected to play an important role in monitoring gamma-ray transients all over the sky. Since the launch on July 27th, 2022, GECAM-C has detected hundreds of gamma-ray bursts, X-ray bursts and solar flares, including the GRB 221009A, which is the brightest GRBs ever detected\cite{GCN32751}, and the second X-ray burst associated with fast radio burst \cite{ATel15682}. Detailed calibration tests and studies presented in this work will help us to better understand the performance of GECAM-C GRD detectors and to analyze the observation data of GECAM-C.

\section*{Acknowledgements}
We thank the staff of the Key Laboratory of Particle Astrophysics, Center for Space Science and National Institute of Metrology who offered great help in the development and calibration of GECAM-C. This research is supported by National Key R\&D Program of China (Grant No. 2021YFA0718500), the Strategic Priority Research Program of Chinese Academy of Sciences (GranNo. XDA15360102), the National Natural Science Foundation of China (Grant No. 12273042).

\doublespacing
\bibliographystyle{unsrt}
\bibliography{main.bib}

\begin{thebibliography}{10}

\bibitem{GW170817}
Benjamin~P Abbott, Rich Abbott, TD~Abbott, Fausto Acernese, Kendall Ackley,
  Carl Adams, Thomas Adams, Paolo Addesso, RX~Adhikari, Vaishali~B Adya, et~al.
\newblock Gw170817: observation of gravitational waves from a binary neutron
  star inspiral.
\newblock {\em Physical review letters}, 119(16):161101, 2017.

\bibitem{GRB170817}
Benjamin~P Abbott, Robert Abbott, TD~Abbott, F~Acernese, K~Ackley, C~Adams,
  T~Adams, P~Addesso, RX~Adhikari, VB~Adya, et~al.
\newblock Gravitational waves and gamma-rays from a binary neutron star merger:
  Gw170817 and grb 170817a.
\newblock {\em The Astrophysical Journal Letters}, 848(2):L13, 2017.

\bibitem{Multi_messenger}
Ligo~Scientific Collaboration, V.~Collaboration, Gbm Fermi, INTERGRAL, and
  Yann~Collab Hello.
\newblock Multi-messenger observations of a binary neutron star merger.
\newblock {\em Astrophysical Journal}, 2017.

\bibitem{GECAM}
X.~Q. Li, X.~Y. Wen, Z.~H. An, C.~Cai, Z.~Chang, G.~Chen, C.~Chen, Y.~Y. Du,
  M.~Gao, and R.~Gao.
\newblock The technology for detection of gamma-ray burst with gecam satellite.
\newblock {\em Radiation Detection Technology and Methods}, (006-001), 2022.

\bibitem{xiao2022ground}
S~Xiao, YQ~Liu, WX~Peng, ZH~An, SL~Xiong, YL~Tuo, K~Gong, P~Zhang, K~Zhang,
  SJ~Zheng, et~al.
\newblock On-ground and on-orbit time calibrations of gecam.
\newblock {\em Monthly Notices of the Royal Astronomical Society},
  511(1):964--971, 2022.

\bibitem{xiao2021}
S~Xiao, SL~Xiong, SN~Zhang, LM~Song, FJ~Lu, Y~Huang, C~Cai, QB~Yi, XY~Song,
  W~Chen, et~al.
\newblock Enhanced localization of transients based on a novel
  cross-correlation method.
\newblock {\em The Astrophysical Journal}, 920(1):43, 2021.

\bibitem{objective0}
Xiong~S L.
\newblock Special topic: Gecam gamma-ray all-sky monitor.
\newblock {\em SCIENTIA SINICA Physica, Mechanica \& Astronomica},
  50(12):129501, 2020.

\bibitem{objective2}
Qi~Luo, Shaolin Xiong, Yan Huang, and Binbin Zhang.
\newblock Ultra-long gamma-ray bursts and ultra-soft gamma-ray bursts.
\newblock {\em SCIENTIA SINICA Physica, Mechanica \& Astronomica},
  50(12):129504, 2020.

\bibitem{objective3}
S.~U. Yang, W.~X. Peng, W.~Chen, S.~L. Xiong, and Y.~Zhu.
\newblock Monitoring and research of high-energy solar flare emissions with
  gecam.
\newblock {\em SCIENTIA SINICA Physica, Mechanica \& Astronomica}, 2020.

\bibitem{Liao2020}
J.~Y. Liao, Q.~Luo, Y.~Zhu, X.~Y. Song, W.~X. Peng, S.~Xiao, L.~I. Gang, and
  S.~L. Xiong.
\newblock The localization method of gecam and simulation analysis.
\newblock {\em SCIENTIA SINICA Physica, Mechanica \& Astronomica},
  50(12):129510, 2020.

\bibitem{zhao2023gecam}
Yi~Zhao, Wang-Chen Xue, Shao-Lin Xiong, Yuan-Hao Wang, Jia-Cong Liu, Qi~Luo,
  Yan-Qiu Zhang, Jian-Chao Sun, Xiao-Yun Zhao, Ce~Cai, et~al.
\newblock Gecam localization of high-energy transients and the systematic
  error.
\newblock {\em The Astrophysical Journal Supplement Series}, 265(1):17, 2023.

\bibitem{GECAMDesign1}
Xinqiao Li, Xiangyang Wen, , Zhenghua An, Yanbing Xu, Xiaohua Liang, Sheng
  Yang, Xilei Sun, Xiaojing Liu, Min Gao, Jinzhou Wang, et~al.
\newblock The gecam and its payload.
\newblock {\em SCIENTIA SINICA Physica, Mechanica \& Astronomica},
  50(12):129508, 2020.

\bibitem{GRDsDesign2}
Z.~H. An, X.~L. Sun, D.~L. Zhang, S.~Yang, X.~Q. Li, X.~Y. Wen, K.~Gong, X.~H.
  Liang, X.~J. Liu, and Y.~Q. Liu.
\newblock The design and performance of grd onboard the gecam satellite.
\newblock {\em Radiation Detection Technology and Methods}, (006-001), 2022.

\bibitem{CPDsDesign}
YB~Xu, XQ~Li, XL~Sun, S~Yang, H~Wang, WX~Peng, XH~Liang, K~Gong, YQ~Liu,
  DY~Guo, et~al.
\newblock The design and performance of charged particle detector onboard the
  gecam mission.
\newblock {\em Radiat Detect Technol Methods}, 6:53--62, 2022.

\bibitem{Zhang2023}
Dali Zhang et~al.
\newblock The performance of sipm-based gamma-ray detector onboard gecam-c.
\newblock In-prepare.

\bibitem{geant4}
Sea Agostinelli, John Allison, K~al Amako, John Apostolakis, H~Araujo, Pedro
  Arce, Makoto Asai, D~Axen, Swagato Banerjee, GJNI Barrand, et~al.
\newblock Geant4—a simulation toolkit.
\newblock {\em Nuclear instruments and methods in physics research section A:
  Accelerators, Spectrometers, Detectors and Associated Equipment},
  506(3):250--303, 2003.

\bibitem{HXCF_1}
X.~Zhou, X.Q. Li, and Y.N. et~al. Xie.
\newblock Introduction to a calibration facility for hard x-ray detectors.
\newblock {\em Experimental Astronomy}, 38:433--441, 2014.

\bibitem{HXCF_2}
Si~Ming Guo, Jin~Jie Wu, and Dong~Jie Hou.
\newblock The development, performances and applications of the monochromatic
  x-rays facilities in (0.218–301) kev at nim, china.
\newblock {\em Nuclear Science and Techniques}, 2021.

\bibitem{2019Ground}
X.~F. Li, C.~Z. Liu, Z.~Chang, Y.~F. Zhang, and X.~P. Qiu.
\newblock Ground-based calibration and characterization of the he detectors for
  insight-hxmt.
\newblock {\em Journal of High Energy Astrophysics}, 24, 2019.

\bibitem{GRID_calibration}
Huaizhong Gao, Dongxin Yang, Jiaxing Wen, Xutao Zheng, Ming Zeng, Jirong Cang,
  Weihe Zeng, Xiaofan Pan, Qimin Zhou, Yihui Liu, et~al.
\newblock On-ground calibrations of the grid-02 gamma-ray detector.
\newblock {\em Experimental Astronomy}, 53(1):103--116, 2022.

\bibitem{gecam_calibration_GRD}
JJ~He, ZH~An, WX~Peng, XQ~Li, SL~Xiong, DL~Zhang, R~Qiao, DY~Guo, C~Cai,
  Z~Chang, et~al.
\newblock Ground-based calibration and characterization of grd of gecam: 8-160
  kev.
\newblock {\em arXiv preprint arXiv:2112.04787}, 2021.

\bibitem{SVOM_calibration_GRM}
X.~Wen, J.~Sun, J.~He, R.~Song, and S.~Zhang.
\newblock Calibration study of the gamma-ray monitor onboard the svom
  satellite.
\newblock {\em Nuclear Instruments and Methods in Physics Research Section A
  Accelerators Spectrometers Detectors and Associated Equipment}, (1):165301,
  2021.

\bibitem{BeamMonitor}
Tao Yu, Siming Guo, Jinjie Wu, Zhenghua An, Xiaoyu Qie, and Kaiyue Guo.
\newblock Research on the temperature characteristics of sipm-based labr3 (ce)
  detectors.
\newblock {\em Journal of the Korean Physical Society}, pages 1--7, 2022.

\bibitem{LEGe}
Liu Haoran, Wu~Jinjie, Liang Juncheng, Chen Fajun, and Li~Zeshu.
\newblock Lege detector intrinsic efficiency calibration for parallel incident
  photons.
\newblock {\em Applied Radiation and Isotopes}, 109:551--554, 2016.

\bibitem{qie2021study}
Xiaoyu Qie, Siming Guo, Jiang He, Zheng Jiang, Tao Yu, Xing Wen, Shiwei Ren,
  Zhenghua An, and Jinjie Wu.
\newblock Study on the performance of sy-01 satellite gamma-ray detector.
\newblock In {\em 2021 IEEE 15th International Conference on Electronic
  Measurement \& Instruments (ICEMI)}, pages 254--258. IEEE, 2021.

\bibitem{Bhat2008Ground}
Bhat, Greiner, Krumrey, V.~D. Horst, Kouveliotou, Paciesas, Bissaldi, Kippen,
  Preece, and Diehl.
\newblock Ground-based calibration and characterization of the fermi gamma-ray
  burst monitor detectors, 2008.

\bibitem{Scintillator_nPR_1}
W.~W. Moses, S.~A. Payne, W.~S Choong, G.~Hull, and B.~W. Reutter.
\newblock Scintillator non-proportionality: Present understanding and future
  challenges.
\newblock {\em IEEE Transactions on Nuclear ence}, 55(3):1049--1053, 2008.

\bibitem{Scintillator_nPR_2}
I.~V. Khodyuk, Jtm~De Haas, and P.~Dorenbos.
\newblock Nonproportional response between 0.1–100 kev energy by means of
  highly monochromatic synchrotron x-rays.
\newblock {\em IEEE Transactions on Nuclear Science}, 57(3):1175--1181, 2010.

\bibitem{Scintillator_nPR_3}
B.~D. Rooney and J.~D. Valentine.
\newblock Benchmarking the compton coincidence technique for measuring electron
  response nonproportionality in inorganic scintillators.
\newblock {\em IEEE Transactions on Nuclear Science}, 43(3):1271--1276, 2002.

\bibitem{Li2020}
X.~Li, X.~Li, Y.~Tan, Y.~Yang, and C.~Li.
\newblock In-flight calibration of the insight-hard x-ray modulation telescope.
\newblock {\em Journal of High Energy Astrophysics}, 27, 2020.

\bibitem{Zheng2022}
C~Zheng, WX~Peng, XB~Li, ZH~An, SL~Xiong, XF~Lan, DL~Zhang, CY~Li, R~Qiao,
  DY~Guo, et~al.
\newblock Electron non-linear light yield of labr3 detector aboard gecam.
\newblock {\em Nuclear Instruments and Methods in Physics Research Section A:
  Accelerators, Spectrometers, Detectors and Associated Equipment},
  1042:167427, 2022.

\bibitem{Li2023}
Chaoyang Li et~al.
\newblock On‑ground calibration of low gain response for gamma‑ray
  detectors onboard the gecam satellite.
\newblock In-prepare.

\bibitem{Fermi_GBM}
E.~Bissaldi, A.~V. Kienlin, G.~Lichti, H.~Steinle, P.~N. Bhat, M.~S. Briggs,
  G.~J. Fishman, A.~S. Hoover, R.~M. Kippen, and M.~Krumrey.
\newblock Ground-based calibration and characterization of the fermi gamma-ray
  burst monitor detectors.
\newblock {\em Experimental Astronomy}, 24(1-3):47--88, 2009.

\bibitem{GCN32751}
J.~C. Liu, Y.~Q. Zhang, S.~L. Xiong, C.~Zheng, C.~W. Wang, W.~C. Xue, et~al.
\newblock Grb 221009a: Hebs detection.
\newblock {\em GRB Coordinates Network}, 32751:1, October 2022.

\bibitem{ATel15682}
C.~W. Wang, S.~L. Xiong, Y.~Q. Zhang, J.~C. Liu, C.~Zheng, W.~C. Xue, w.J. Tan,
  et~al.
\newblock Ecam and hebs detection of a short x-ray burst from sgr j1935+2154
  associated with radio burst.
\newblock {\em The Astronomer's Telegram}, 15682:1, October 2022.

\end{thebibliography}

\end{document}